\pdfoutput=1

\documentclass[11pt]{article}

\usepackage[final]{acl}

\usepackage{times}
\usepackage{latexsym}
\usepackage{CJKutf8}

\usepackage[T1]{fontenc}

\usepackage[utf8]{inputenc}

\usepackage{microtype}

\usepackage{inconsolata}

\usepackage{graphicx}

\usepackage{multirow}  
\usepackage{booktabs} 
\usepackage{array}
\usepackage{tcolorbox}
\usepackage{adjustbox}
\usepackage{longtable}
\usepackage{amsmath}
\usepackage{soul} 
\usepackage[caption=false]{subfig}
\usepackage{colortbl} 
\usepackage{xcolor}   

\title{MiLQ: Benchmarking IR Models for Bilingual Web Search \\
with Mixed Language Queries}

\author{
 \textbf{Jonghwi Kim\textsuperscript{1}},
 \textbf{Deokhyung Kang\textsuperscript{1}},
 \textbf{Seonjeong Hwang\textsuperscript{1}},
 \\
 \textbf{Yunsu Kim\textsuperscript{3}}\thanks{This work was done when the author was at aiXplain},
 \textbf{Jungseul Ok\textsuperscript{1,2}},
 \textbf{Gary Geunbae Lee\textsuperscript{1,2}},
\\
 \textsuperscript{1} Graduate School of Artificial Intelligence, POSTECH, Republic of Korea ,
 \\
 \textsuperscript{2} Department of Computer Science and Engineering, POSTECH, Republic of Korea,
 \\
 \textsuperscript{3} Lilt, Inc.
 \\
 \small{\texttt{\{jonghwi.kim, deokhk, seonjeongh, jungseul.ok, gblee\}@postech.ac.kr}}, 
 \small{\texttt{\href{mailto:email@domain}{yunsu.kim@lilt.com}}}
}

\definecolor{scoreRed01}{RGB}{255,240,240} 
\definecolor{scoreRed02}{RGB}{255,235,235} 
\definecolor{scoreRed03}{RGB}{255,230,230} 
\definecolor{scoreRed04}{RGB}{255,225,225} 
\definecolor{scoreRed05}{RGB}{255,220,220} 
\definecolor{scoreRed06}{RGB}{255,215,215} 
\definecolor{scoreRed07}{RGB}{255,210,210} 
\definecolor{scoreRed08}{RGB}{255,205,205} 
\definecolor{scoreRed09}{RGB}{255,200,200} 
\definecolor{scoreRed10}{RGB}{255,195,195} 
\definecolor{scoreRed11}{RGB}{255,190,190} 
\definecolor{scoreRed12}{RGB}{255,180,180} 
\definecolor{scoreRed13}{RGB}{255,170,170} 
\definecolor{scoreRed14}{RGB}{255,160,160} 
\definecolor{scoreRed15}{RGB}{255,150,150} 
\definecolor{scoreRed16}{RGB}{255,140,140} 
\definecolor{scoreRed17}{RGB}{255,130,130} 
\definecolor{scoreRed18}{RGB}{255,120,120} 
\definecolor{scoreRed19}{RGB}{255,110,110} 
\definecolor{scoreRed20}{RGB}{255,100,100} 

\definecolor{scoreBlue00}{RGB}{230,240,255} 
\definecolor{scoreBlue01}{RGB}{224,235,252} 
\definecolor{scoreBlue02}{RGB}{218,229,250} 
\definecolor{scoreBlue03}{RGB}{211,224,247} 
\definecolor{scoreBlue04}{RGB}{205,218,245} 
\definecolor{scoreBlue05}{RGB}{199,213,242} 
\definecolor{scoreBlue06}{RGB}{193,207,240} 
\definecolor{scoreBlue07}{RGB}{187,202,237} 
\definecolor{scoreBlue08}{RGB}{181,196,235} 
\definecolor{scoreBlue09}{RGB}{174,191,232} 
\definecolor{scoreBlue10}{RGB}{168,185,230} 
\definecolor{scoreBlue11}{RGB}{162,180,227} 
\definecolor{scoreBlue12}{RGB}{156,174,225} 
\definecolor{scoreBlue13}{RGB}{150,169,222} 
\definecolor{scoreBlue14}{RGB}{143,163,220} 
\definecolor{scoreBlue15}{RGB}{137,158,217} 
\definecolor{scoreBlue16}{RGB}{131,152,215} 
\definecolor{scoreBlue17}{RGB}{125,147,212} 
\definecolor{scoreBlue18}{RGB}{119,141,210} 
\definecolor{scoreBlue19}{RGB}{112,136,207} 
\definecolor{scoreBlue20}{RGB}{106,130,205} 
\definecolor{scoreBlue21}{RGB}{100,125,202} 
\definecolor{scoreBlue22}{RGB}{94,119,200}  
\definecolor{scoreBlue23}{RGB}{88,114,197}  
\definecolor{scoreBlue24}{RGB}{81,108,195}  
\definecolor{scoreBlue25}{RGB}{75,103,192}  
\definecolor{scoreBlue26}{RGB}{69,97,190}   
\definecolor{scoreBlue27}{RGB}{63,92,187}   
\definecolor{scoreBlue28}{RGB}{57,86,185}   
\definecolor{scoreBlue29}{RGB}{50,80,180}   

\begin{document}
\maketitle

\begin{abstract}

Despite bilingual speakers frequently using mixed-language queries in web searches, Information Retrieval (IR) research on them remains scarce.
To address this, we introduce \textbf{MiLQ}, \underline{\textbf{Mi}}xed-\underline{\textbf{L}}anguage \underline{\textbf{Q}}uery test set, the first public benchmark of mixed-language queries, qualified as realistic and relatively preferred.
Experiments show that multilingual IR models perform moderately on MiLQ and inconsistently across native, English, and mixed-language queries, also suggesting code-switched training data's potential for robust IR models handling such queries.
Meanwhile, intentional English mixing in queries proves an effective strategy for bilinguals searching English documents, which our analysis attributes to enhanced token matching compared to native queries.\footnote{The code for this work are available at : \href{https://github.com/jonghwi-kim/milq}{https://github.com/jonghwi-kim/milq}}


\end{abstract}

\section{Introduction}
Code-switching\footnote{In this study, code-switching, mixed-language, and code-mixing are used synonymously.}, where bilingual speakers alternate languages within a context, is a prevalent linguistic behavior in multilingual communities \cite{auer1999codeswitching, gardner2009code, auer2013code}.
This phenomenon extends to Human-Computer Interaction (HCI), especially via AI agents like ChatGPT \cite{openai2023chatgpt}, where understanding mixed-language input critically affects their perceived reliability by bilingual users \cite{bawa2020multilingual, choi2023toward}.
Information Retrieval (IR) systems also face the challenge of effectively handling such mixed-language queries \cite{sitaram2019survey}.

Meanwhile, recent IR research has expanded beyond Monolingual IR (\textit{MonoIR}) settings to diverse multilingual settings. 
The benchmarks \cite{asai-etal-2021-xor, lawrie2023hc3, lawrie2023overview, soboroff2023better, adeyemi2024ciral, litschko2025cross} are widely utilized, representing diverse language scenarios.
However, research on mixed-language queries remains sparse and outdated \cite{fung1999mixed, gupta2014query, sequiera2015overview}, with no publicly available benchmark.

\begin{figure}[t!]
\centering
\includegraphics[width=0.49\textwidth]{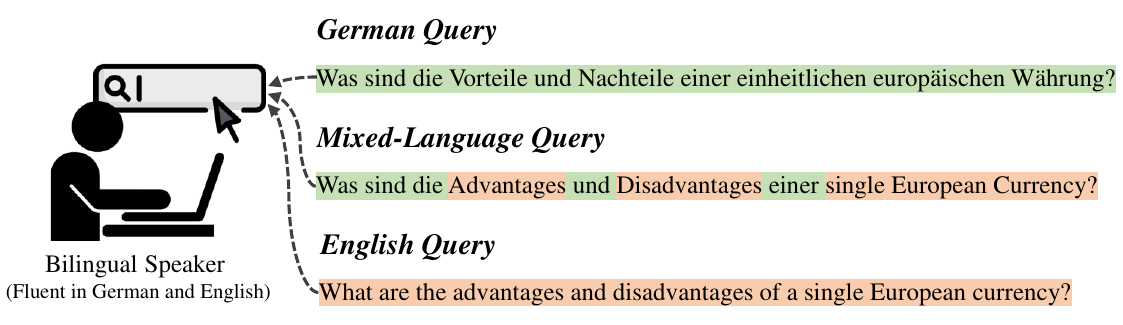}
\caption{
\small{Illustration of a bilingual user freely using German, English, and mixed-language queries. German elements are in \colorbox[rgb]{0.7,0.9,0.7}{green}, and English in \colorbox[rgb]{1.0,0.8,0.7}{orange}.}\vspace{-1pt}}
\label{fig:different_query_languages}
\end{figure}

To address these gaps, we introduce \textbf{MiLQ}, the first \textbf{\underline{Mi}}xed-\textbf{\underline{L}}anguage \textbf{\underline{Q}}uery benchmark created by actual bilingual users (Figure\ref{fig:different_query_languages}).
Using MiLQ, we explore three main research questions: 
\textbf{(RQ1)} How realistic are our mixed-language queries, and which query language do bilingual users prefer?
\textbf{(RQ2)} How well do existing multilingual IR models perform in Mixed-language Query Information Retrieval (\textit{MQIR})?
\textbf{(RQ3)} Is the behavior of intentionally mixing English terms into query, noted in HCI studies \cite{fu2017query, fu2019mixed}, an effective strategy?

The main contributions of our work are:
\begin{itemize}
    \item We introduce \textbf{MiLQ}\footnote{
    \href{https://catalogue.elra.info/en-us/repository/browse/ELRA-E0047/}{https://catalogue.elra.info/en-us/repository/browse/ELRA-E0047/}
    }, the first public benchmark of mixed-language queries, qualified as realistic and relatively preferred by bilinguals.
    \item We provide a comprehensive performance analysis of multilingual IR models on \textbf{MiLQ}, establishing initial baselines for \textit{MQIR}.
    \item We show intentionally mixed-language queries are effective for English document retrieval across diverse methods, providing token-level analysis of their rationale.
\end{itemize}




\begin{table*}[htbp]
  \centering
  \footnotesize 
  \resizebox{0.95\textwidth}{!}{
  \begin{tabular}{c|c|c||c|cc|ccc||c|cc|ccc}
    \toprule
    \textbf{Retrieval} & \textbf{Native} & \textbf{Num.} & \multicolumn{6}{c||}{\fontsize{10}{11}\selectfont{\textbf{Title Query}}} & \multicolumn{6}{c}{\fontsize{10}{11}\selectfont{\textbf{Description Query}}} \\
    \textbf{Scenario} & \textbf{Lang} & \textbf{of} & \textbf{CMI} & \multicolumn{2}{c|}{\textbf{GPT-Eval}} & \multicolumn{3}{c||}{\textbf{Human-Eval}} & \textbf{CMI} & \multicolumn{2}{c|}{\textbf{GPT-Eval}} & \multicolumn{3}{c}{\textbf{Human-Eval}} \\
    \textbf{(Q$\rightarrow$D)} & \textbf{(XX)} & \textbf{Query} & \textbf{(XX$\rightarrow$MiLQ)} & \textbf{Acc.} & \textbf{Flu.} & \textbf{Acc.} & \textbf{Flu.} & \textbf{Real.} & \textbf{(XX$\rightarrow$MiLQ)} & \textbf{Acc.} & \textbf{Flu.} & \textbf{Acc.} & \textbf{Flu.} & \textbf{Real.} \\
    \midrule
    \multirow{5}{*}{\textbf{Mixed$\rightarrow$EN}} & SW & 151 & 8.4$\rightarrow$38.6 & \cellcolor{scoreRed14}{2.35} & \cellcolor{scoreRed14}{2.39} & \cellcolor{scoreRed19}{2.83} & \cellcolor{scoreRed17}{2.65} & \cellcolor{scoreRed17}{2.66} & 5.6$\rightarrow$30.7 & \cellcolor{scoreRed19}{2.83} & \cellcolor{scoreRed15}{2.44} & \cellcolor{scoreRed19}{2.83} & \cellcolor{scoreRed17}{2.62} & \cellcolor{scoreRed17}{2.61} \\
    & SO & 151 & 16.2$\rightarrow$59.6 & \cellcolor{scoreRed14}{2.38} & \cellcolor{scoreRed14}{2.34} & \cellcolor{scoreRed18}{2.73} & \cellcolor{scoreRed16}{2.58} & \cellcolor{scoreRed20}{2.95} & 5.4$\rightarrow$36.1 & \cellcolor{scoreRed18}{2.76} & \cellcolor{scoreRed14}{2.34} & \cellcolor{scoreRed17}{2.63} & \cellcolor{scoreRed16}{2.51} & \cellcolor{scoreRed18}{2.77} \\
    & FI & 151 & 7.3$\rightarrow$40.2 & \cellcolor{scoreRed15}{2.48} & \cellcolor{scoreRed16}{2.52} & \cellcolor{scoreRed18}{2.79} & \cellcolor{scoreRed18}{2.70} & \cellcolor{scoreRed16}{2.59} & 2.2$\rightarrow$45.3 & \cellcolor{scoreRed17}{2.63} & \cellcolor{scoreRed12}{2.15} & \cellcolor{scoreRed17}{2.63} & \cellcolor{scoreRed15}{2.44} & \cellcolor{scoreRed13}{2.28} \\
    & DE & 151 & 9.1$\rightarrow$61.8 & \cellcolor{scoreRed17}{2.67} & \cellcolor{scoreRed17}{2.68} & \cellcolor{scoreRed17}{2.61} & \cellcolor{scoreRed16}{2.50} & \cellcolor{scoreRed13}{2.21} & 2.1$\rightarrow$41.1 & \cellcolor{scoreRed16}{2.55} & \cellcolor{scoreRed12}{2.11} & \cellcolor{scoreRed15}{2.43} & \cellcolor{scoreRed12}{2.15} & \cellcolor{scoreRed09}{1.80} \\
    & FR & 151 & 5.7$\rightarrow$35.0 & \cellcolor{scoreRed16}{2.52} & \cellcolor{scoreRed16}{2.55} & \cellcolor{scoreRed19}{2.84} & \cellcolor{scoreRed16}{2.51} & \cellcolor{scoreRed14}{2.31} & 2.3$\rightarrow$32.9 & \cellcolor{scoreRed19}{2.80} & \cellcolor{scoreRed14}{2.30} & \cellcolor{scoreRed19}{2.84} & \cellcolor{scoreRed16}{2.51} & \cellcolor{scoreRed14}{2.31} \\
    \midrule
    \multirow{3}{*}{\textbf{Mixed$\rightarrow$XX}} & ZH & 47 & 0.3$\rightarrow$13.7 & \cellcolor{scoreRed19}{2.85} & \cellcolor{scoreRed19}{2.85} & \cellcolor{scoreRed18}{2.79} & \cellcolor{scoreRed18}{2.79} & \cellcolor{scoreRed17}{2.64} & 2.4$\rightarrow$9.0 & \cellcolor{scoreRed19}{2.89} & \cellcolor{scoreRed20}{2.91} & \cellcolor{scoreRed17}{2.65} & \cellcolor{scoreRed18}{2.70} & \cellcolor{scoreRed16}{2.50} \\
    & FA & 45 & 2.2$\rightarrow$15.0 & \cellcolor{scoreRed20}{2.98} & \cellcolor{scoreRed20}{2.98} & \cellcolor{scoreRed19}{2.87} & \cellcolor{scoreRed19}{2.82} & \cellcolor{scoreRed17}{2.64} & 0.1$\rightarrow$5.6 & \cellcolor{scoreRed20}{3.00} & \cellcolor{scoreRed20}{2.93} & \cellcolor{scoreRed20}{2.91} & \cellcolor{scoreRed19}{2.81} & \cellcolor{scoreRed17}{2.68} \\
    & RU & 44 & 0.0$\rightarrow$51.7 & \cellcolor{scoreRed19}{2.89} & \cellcolor{scoreRed16}{2.50} & \cellcolor{scoreRed18}{2.72} & \cellcolor{scoreRed14}{2.30} & \cellcolor{scoreRed12}{2.16} & 0.6$\rightarrow$51.7 & \cellcolor{scoreRed20}{2.93} & \cellcolor{scoreRed15}{2.45} & \cellcolor{scoreRed18}{2.73} & \cellcolor{scoreRed12}{2.16} & \cellcolor{scoreRed12}{2.14} \\
    \midrule
    \multicolumn{2}{c|}{\textbf{Average}} & \textbf{111.4} & \textbf{6.2$\rightarrow$39.5} & \cellcolor{scoreRed17}{\textbf{2.64}} & \cellcolor{scoreRed17}{\textbf{2.60}} & \cellcolor{scoreRed18}{\textbf{2.78}} & \cellcolor{scoreRed16}{\textbf{2.59}} & \cellcolor{scoreRed16}{\textbf{2.57}} & \textbf{5.0$\rightarrow$31.6} & \cellcolor{scoreRed19}{\textbf{2.80}} & \cellcolor{scoreRed15}{\textbf{2.45}} & \cellcolor{scoreRed18}{\textbf{2.76}} & \cellcolor{scoreRed15}{\textbf{2.45}} & \cellcolor{scoreRed15}{\textbf{2.46}} \\
    \bottomrule
  \end{tabular}}
  \caption{\footnotesize{Quality measurements for \textbf{MiLQ} (Title \& Description queries). Code-Mixing Index (CMI) is on a 0-100 scale (Original Query CMI $\rightarrow$ Mixed-language Query (\textbf{MiLQ}) CMI). For GPT-Eval (Accuracy [Acc.] \& Fluency [Flu.]) and Human-Evaluation (Acc. \& Flu. \& Realism [Real.]), both on a 1-3 scale, cell backgrounds are colored in a fine-grained red gradient from \colorbox{scoreRed01}{lightest red (scores $\approx$1.0)} to \colorbox{scoreRed20}{darkest red (scores $\approx$3.0)}. The 'Average' row is \textbf{bolded}. "XX" denotes the native language.}}
  \label{tab:stat_milq}
\end{table*}

\section{MiLQ: \textbf{Mi}xed-\textbf{L}anguage \textbf{Q}uery test set}

\paragraph{Data Construction}

We started with queries from two Cross-Language IR (\textit{CLIR}) benchmarks: CLEF \cite{braschler2003clef} and NeuCLIR22 \cite{lawrie2023overview}, addressing native-to-English and English-to-native retrieval, respectively.
These were selected to ensure diverse language scenarios while maintaining quality, based on three criteria: (1) availability of parallel English and native-language queries, (2) widespread use for performance comparison, and (3) budgetary feasibility.
Both follow the TREC format~\cite{voorhees2005trec}, including short Title and longer Description queries, for which we created mixed-language versions.

Bilingual speakers, experienced in both languages and mixed-language search, crafted natural mixed-language queries from original English and native query pairs, while preserving the original search intent.
To reflect realistic code-switching patterns, we adopt Matrix Language Frame theory \cite{myers1997duelling} and follow prior studies \cite{fu2017query, fu2019mixed, yong2023prompting, winata2023decades} that describe common code-switching as featuring native language as the grammar-governing matrix and English language as embedded.
Accordingly, annotators integrated English terms into the native language structure only when conceptually necessary and linguistically sound.
Annotation guidelines are in Appendix \ref{sec:appendix_Annotation_Details}, and MiLQ samples are in Appendix \ref{sec:appendix_title_example} (Figures \ref{fig:milq_examples_title}, \ref{fig:milq_examples_description}).

\begin{figure*}[ht!]
\centering
\includegraphics[width=0.95\textwidth]{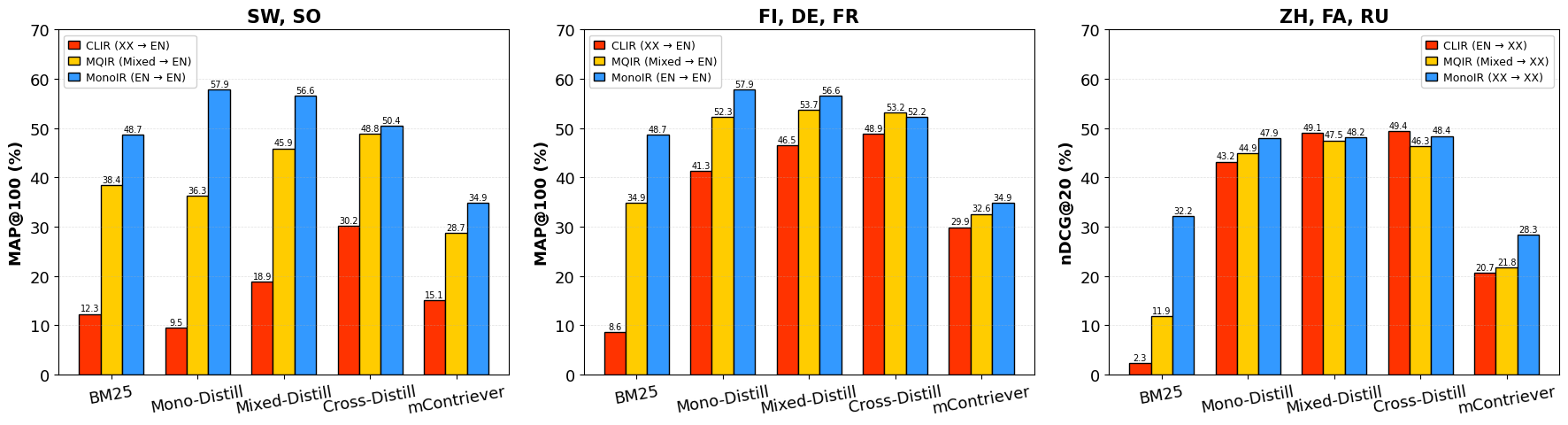} 
\caption{\small{Performance of retrieval models across \colorbox[rgb]{1,0.4,0}{\textit{CLIR}}, \colorbox[rgb]{1,0.8,0}{\textit{MQIR} (\textbf{MiLQ})}, and  \colorbox[rgb]{0.4,0.6,1}{\textit{MonoIR}} scenarios. Results are averaged by language group: low-resource (SW, SO; MAP@100) [left], high-resource (FI, DE, FR; MAP@100) [middle], and diverse document language (ZH, FA, RU; nDCG@20) [right]. Models include \textbf{BM25}, specialized multi-vector dense retrievers (\textbf{Mono-, Mixed-, Cross-Distill}), and \textbf{mContriever}. See Appendix \ref{sec:appendix_individual_performance} for per-language details.}
}
\label{fig:main_result_1}
\end{figure*}

\paragraph{Quality Measurement and Analysis}
We measured MiLQ's quality considering its language mixing, meaning preservation, naturalness, and realism (Table \ref{tab:stat_milq}).
First, for language mixing, we used the \textbf{Code-Mixing Index (CMI)} \cite{das2014identifying} (0-100 scale, higher=more mixing; Appendix \ref{sec:cmi}). Average CMI increased from 6.2 to 39.5 (Title) and 5.0 to 31.6 (Description), showing substantially more mixing than originals.
Next, \textbf{GPT-Eval} (GPT-4o) using \citeauthor{kuwanto2024linguistics}'s framework (high human alignment, Kendall’s Tau > 0.5) assessed MiLQ (1-3 scale; rubrics in Appendix \ref{sec:gpt_eval_rubrics}) for \textbf{Accuracy (Acc.)} (meaning preservation, correct term use) and \textbf{Fluency (Flu.)} (naturalness, readability, seamlessness). 
MiLQ achieved strong average GPT-Eval scores: Acc. 2.64 / Flu. 2.60 (Title) and Acc. 2.80 / Flu. 2.45 (Description).
Lastly, for \textbf{Human-Eval}, three bilingual annotators per query assessed MiLQ on a 1-3 scale (detailed guidelines in Appendix \ref{sec:human_eval_guidelines}). This evaluation covered \textbf{Accuracy (Acc.)} and \textbf{Fluency (Flu.)}, using criteria consistent with GPT-Eval, and an additional \textbf{Realism (Real.)}. 
Realism specifically assessed how naturally bilingual speakers might use the given mixed-language query in real search scenarios. 
Human evaluators rated MiLQ highly, with average scores: Acc. 2.78 / Flu. 2.59 / Real. 2.57 (Title) and Acc. 2.76 / Flu. 2.45 / Real. 2.46 (Description).
These consistently high scores in all metrics affirm the quality and reliability of MiLQ.

\begin{table}[htbp]
  \centering
  \resizebox{0.95\columnwidth}{!}{%
  \begin{tabular}{@{}c | cccc | cccc@{}}
    \toprule
    \textbf{Native} & \multicolumn{4}{c|}{\textbf{Title}} & \multicolumn{4}{c}{\textbf{Description}} \\
    \cmidrule(lr){2-5} \cmidrule(lr){6-9}
    \textbf{Lang.} & \textbf{Agr.(\%)} & \textbf{Nat.} & \textbf{Mix.} & \textbf{Eng.} & \textbf{Agr.(\%)} & \textbf{Nat.} & \textbf{Mix.} & \textbf{Eng.} \\
    \midrule
    SW & 78.8 & \cellcolor{scoreBlue04}{0.44} & \cellcolor{scoreBlue08}{0.82} & \cellcolor{scoreBlue20}{2.01} & 88.7 & \cellcolor{scoreBlue03}{0.30} & \cellcolor{scoreBlue10}{1.05} & \cellcolor{scoreBlue16}{1.68} \\
    SO & 98.0 & \cellcolor{scoreBlue07}{0.73} & \cellcolor{scoreBlue22}{2.26} & \cellcolor{scoreBlue00}{0.01} & 100.0 & \cellcolor{scoreBlue01}{0.19} & \cellcolor{scoreBlue27}{2.76} & \cellcolor{scoreBlue00}{0.05} \\
    FI & 70.9 & \cellcolor{scoreBlue16}{1.60} & \cellcolor{scoreBlue11}{1.16} & \cellcolor{scoreBlue12}{1.26} & 81.5 & \cellcolor{scoreBlue18}{1.84} & \cellcolor{scoreBlue03}{0.34} & \cellcolor{scoreBlue11}{1.16} \\ 
    DE & 53.0 & \cellcolor{scoreBlue11}{1.11} & \cellcolor{scoreBlue04}{0.46} & \cellcolor{scoreBlue13}{1.38} & 77.5 & \cellcolor{scoreBlue07}{0.79} & \cellcolor{scoreBlue03}{0.38} & \cellcolor{scoreBlue18}{1.85} \\
    FR & 69.5 & \cellcolor{scoreBlue10}{1.02} & \cellcolor{scoreBlue18}{1.86} & \cellcolor{scoreBlue04}{0.47} & 78.8 & \cellcolor{scoreBlue08}{0.82} & \cellcolor{scoreBlue18}{1.89} & \cellcolor{scoreBlue03}{0.38} \\
    ZH & 53.2 & \cellcolor{scoreBlue14}{1.40} & \cellcolor{scoreBlue09}{0.92} & \cellcolor{scoreBlue10}{1.08} & 70.2 & \cellcolor{scoreBlue08}{0.88} & \cellcolor{scoreBlue09}{0.94} & \cellcolor{scoreBlue14}{1.48} \\
    FA & 60.0 & \cellcolor{scoreBlue07}{0.78} & \cellcolor{scoreBlue20}{2.00} & \cellcolor{scoreBlue03}{0.37} & 66.7 & \cellcolor{scoreBlue07}{0.73} & \cellcolor{scoreBlue20}{2.00} & \cellcolor{scoreBlue04}{0.47} \\
    RU & 72.7 & \cellcolor{scoreBlue05}{0.59} & \cellcolor{scoreBlue16}{1.69} & \cellcolor{scoreBlue09}{0.91} & 75.0 & \cellcolor{scoreBlue05}{0.52} & \cellcolor{scoreBlue17}{1.76} & \cellcolor{scoreBlue07}{0.79} \\
    \midrule
    AVG. & 69.3 & \cellcolor{scoreBlue08}{0.87} & \cellcolor{scoreBlue14}{1.43} & \cellcolor{scoreBlue08}{0.89} & 79.6 & \cellcolor{scoreBlue06}{0.60} & \cellcolor{scoreBlue15}{1.54} & \cellcolor{scoreBlue09}{0.96} \\
    \bottomrule
  \end{tabular}%
  }
  \caption{\footnotesize{User preference for Native (Nat.), Mixed-language (Mix.), and English (Eng.) queries.  Agr.(\%): Percentage of queries where a majority (2+ of 3) of annotators agreed on preferred query type(s). Nat./Mix./Eng. values represent average annotator votes (0-3) for each type. Background color intensity indicates preference strength.}}
  \label{tab:user_preference}
\end{table}

To investigate user preferences for Native (Nat.), Mixed-language (Mix.), and English (Eng.) query formulations, we asked annotators to select their preferred formulation(s), allowing for multiple selections.
For robust assessment, Table \ref{tab:user_preference} presents results for queries in which a majority of annotators (2+ of 3) agreed on their preferred formulation. The scores for each formulation type (0-3) represent the average number of annotators who selected that type as preferred.
Overall, Mix. received the highest average scores, with 1.43 for Title and 1.54 for Description queries, outperforming Nat. and Eng. formulations.
However, the degree of preference varied across languages. Notably, Somali (SO) exhibited the strongest preference for mixed-language (e.g., Title: 2.26, Description: 2.76).
To uncover the reasons for such variations, we conducted interviews with annotators in all languages. 
These discussions revealed that Somali speakers frequently code-switch, primarily using English to express modern concepts due to Somali's limited contemporary vocabulary—findings aligned with prior literature \cite{andrzejewski1979language, andrzejewski1978development, kapchits2019somali}. 
Further interview insights, including common themes on mixed-language query usage, are provided in Appendix~\ref{sec:appendix_preference_discussion}.

In summary, this section addressed \textbf{(RQ1)}, confirming that MiLQ is perceived as highly realistic and that bilingual users prefer mixed-language query formulations.
Additional details of MiLQ are in Appendix \ref{sec:appendix_mixed_pos}.



\section{Experimental Setup}

This section details our experimental setup, designed to evaluate various multilingual IR models on mixed-language queries using \textbf{MiLQ}.

\paragraph{Test Scenarios \& Dataset}
We evaluate three retrieval scenarios: \textit{MQIR} (\textbf{MiLQ}) (Mixed$\rightarrow$XX), \textit{MonoIR} (XX$\rightarrow$XX), and \textit{CLIR} (XX$\rightarrow$YY). 
Document collections include NeuCLIR22\footnote{\href{https://catalog.data.gov/dataset/2022-neuclir-dataset}{https://catalog.data.gov/dataset/2022-neuclir-dataset}}~\cite{lawrie2023overview} and CLEF00-03\footnote{\href{https://catalogue.elra.info/en-us/repository/browse/ELRA-E0008/}{https://catalogue.elra.info/en-us/repository/browse/ELRA-E0008/}    } ~\cite{braschler2000clef, 10.1007/3-540-45691-0_2, braschler2002clef, braschler2003clef} (statistics in Appendix \ref{sec:appendixA.2}). Following prior works \cite{huang2023improving, yang2024translate}, queries are concatenations of Title and Description, with MAP@100 and nDCG@20 serving as the primary metrics (detailed in Appendix \ref{sec:metric}).

\paragraph{Retrieval Models} 
To create retrieval models specialized for distinct language scenarios, we developed three ColBERT-based~\cite{khattab2020colbert} dense retrievers: \textbf{Mono-Distill}, \textbf{Cross-Distill}, and \textbf{Mixed-Distill}. 
Based on a multilingual pretrained language model, these models are trained via Knowledge Distillation (KD) adapting Translate-Distill strategy~\cite{yang2024translate} where English IR training data is translated into target languages.
Thus, their specialization for each scenario arises solely from the training data used.
\textbf{Mono-Distill} is trained for \textit{MonoIR} (e.g., XX$\rightarrow$XX or EN$\rightarrow$EN) with monolingual query-document pairs (original MSMARCO ~\cite{nguyen2016ms} or translated version).
\textbf{Cross-Distill} is trained for \textit{CLIR} (e.g., XX$\rightarrow$EN or EN$\rightarrow$XX) with cross-lingual query-document pairs derived from MSMARCO.
\textbf{Mixed-Distill} is trained for \textit{MQIR} (e.g., Mixed$\rightarrow$EN or Mixed$\rightarrow$XX) with artificially code-switched query-document pairs, generated via bilingual lexicon \cite{kamholz2014panlex, conneau2017word} without translation.

We also include the following baselines:
\textbf{mContriever} \cite{izacardunsupervised} serves as a multilingual single vector dense retriever pre-trained for broad language coverage.
\textbf{BM25} \cite{robertson2009probabilistic} is a standard sparse lexical matching retriever.
\textbf{Translate-Test} first translates queries into the document's language via Neural Machine Translation (NMT), then applies BM25 or Mono-Distill for retrieval. 
Detailed implementation specifics for all models are in Appendix \ref{sec:appendix_implementation_details}.

\section{Results and Analysis}

\paragraph{Main Results}


In response to \textbf{(RQ2)}, MiLQ (\textit{MQIR} in Figure \ref{fig:main_result_1}) shows that multilingual IR models like Mono-Distill and Cross-Distill achieve moderate performance in \textit{MQIR}, performing between their \textit{MonoIR} and \textit{CLIR} performance.
This pattern, also observed with the lexical-based BM25, is attributable to \textit{MQIR}'s intermediate level of lexical cues compared to \textit{MonoIR} and \textit{CLIR} settings.

Further observations underscore specialization's limitations.
For instance, Mono-Distill (\textit{MonoIR}-optimized) outperformed Cross-Distill (\textit{CLIR}-optimized) in \textit{MonoIR} settings, and vice-versa.
Additionally, mContriever consistently trails specialized models. 
Notably, Mixed-Distill trained with artificial code-switched text shows well-balanced performance, often outperforming Cross-Distill in \textit{MonoIR} and Mono-Distill in \textit{CLIR/MQIR}. 
This highlights  potential benefits of using mixed-language queries in training for a robust bilingual IR system—a core challenge MiLQ addresses: developing a single robust IR model for bilingual users freely querying in native, English or mixed language.
To better harness this potential of code-switched training data explored in prior studies \cite{litschko2023boosting, liu2025impact}, future work could explore advanced methods, like multilingual LLMs, beyond simple lexicon augmentation.

Regarding \textbf{(RQ3)}, intentionally using mixed-language queries offers context-dependent benefits.
While native queries are optimal for retrieving native content (\textit{MonoIR}, XX$\rightarrow$XX), mixed-language queries (\textit{MQIR}, Mixed$\rightarrow$EN) prove superior to native ones (\textit{CLIR}, XX$\rightarrow$EN) when bilinguals searching English content, thus offering a clear strategic advantage.
Notably, in low-resource \textit{MQIR} for English document retrieval (Figure \ref{fig:main_result_1}, left), BM25 outperforms neural models like mContriever and Mono-Distill.
Consequently, for low-resource languages where neural models struggle with native queries, mixed-language queries with BM25 present a more effective IR system.



\begin{table}[h]
    \centering
    \setlength{\tabcolsep}{4pt}
    \resizebox{1.0\columnwidth}{!}{
    \begin{tabular}{lcc|cc|c}
        \toprule
        \multirow{2}{*}{\textbf{Method}} & \multicolumn{2}{c|}{\cellcolor{blue!10}\textbf{Low Resource}} & \multicolumn{2}{c|}{\cellcolor{red!10}\textbf{High Resource}} & \multirow{2}{*}{\textbf{\textit{MonoIR}}} \\
        \cmidrule(lr){2-3} \cmidrule(lr){4-5}
        & \textbf{\textit{CLIR}} & \textbf{\textit{MQIR}} & \textbf{\textit{CLIR}} & \textbf{\textit{MQIR}} &  \\
        \midrule
        BM25 & 12.35 & 38.35 & 8.56 & 34.92 & \multirow{2}{*}{48.71} \\
        NMT$\rightarrow$BM25 & 41.07 & 48.10 & 46.01 & 47.08 & \\
        \midrule
        Mono-Distill & 9.53 & 36.34 & 41.32 & 52.26 & \multirow{2}{*}{57.93} \\
        NMT$\rightarrow$Mono-Distill & 50.14 & 56.92 & 56.25 & 56.78 &  \\
        \midrule
    \end{tabular}
    }
    \caption{\footnotesize{Performance of BM25 and Mono-Distill before and after applying NMT. The metric used is MAP@100 (\%).}}
    \label{tab:performance_bef_aft_NMT}
\end{table}

\paragraph{Effectiveness of Translate-Test}
Translate-Test, applying NMT at test time, is widely used in \textit{CLIR} \cite{nair2022transfer}.
We evaluated its effectiveness for English document retrieval (XX or Mixed $\rightarrow$ EN), projecting native and mixed-language into English. 
Table~\ref{tab:performance_bef_aft_NMT} shows introducing NMT for both query types consistently improved performance, bringing them closer to the \textit{MonoIR} scenario.
Notably, NMT on mixed-language queries (Mixed$\rightarrow$EN) surpassed NMT on native queries (XX$\rightarrow$EN).
This suggests English terms in mixed queries aid translation, making NMT on these intentionally mixed queries (relevant to \textbf{RQ3}) more effective.
However, current research on Code-Switching Translation~\cite{huzaifah2024evaluating} has been limited to specific language pairs, underscoring the need for tailored NMT models to better support \textit{MQIR}.

\paragraph{Token-Level Analysis for MQIR}

\begin{figure}[ht!]
\centering
\includegraphics[width=0.48\textwidth]{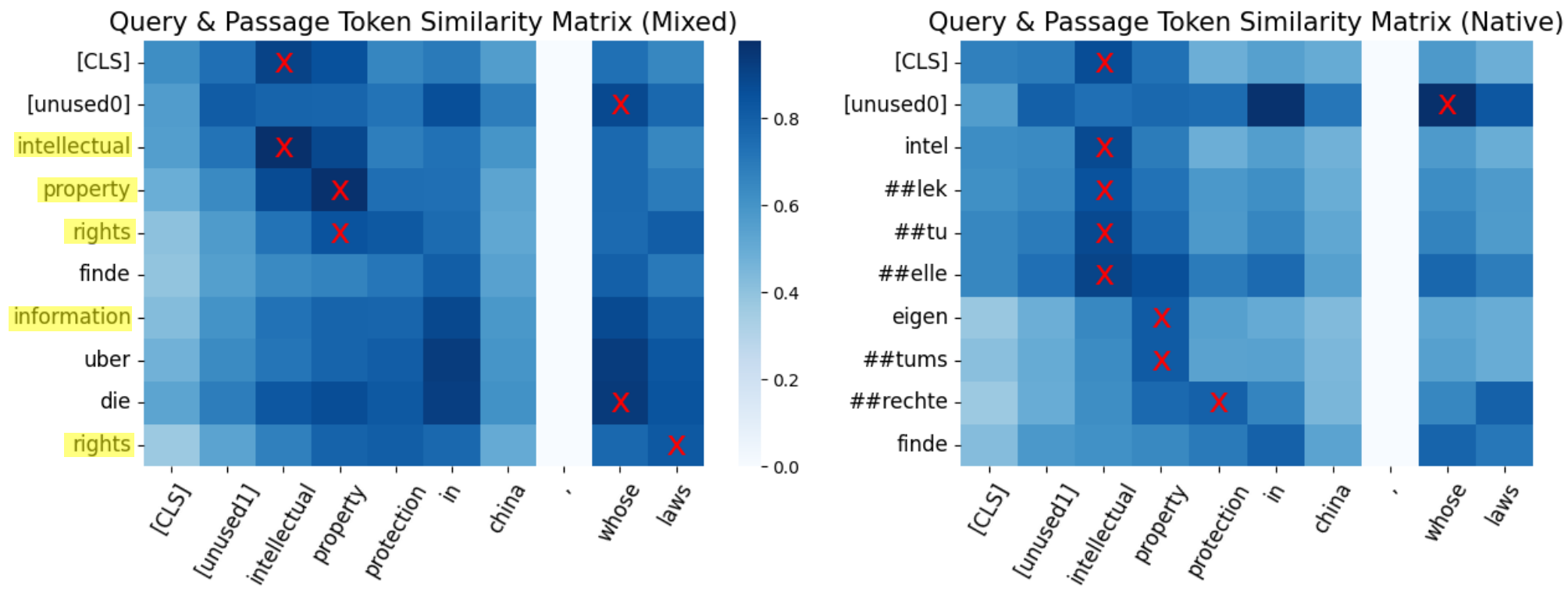} 
\caption{\footnotesize{Token-level similarity matrices from Cross-Distill for German and mixed-language queries on ground truth passage. The y-axis shows tokenized queries (mixed-language left, native right), and the x-axis represents the tokenized English passage. MaxSim tokens are marked by \textcolor{red}{\textbf{$\times$}}, and the code-switched parts are \colorbox[rgb]{1.0, 1.0, 0.3}{highlighted}.}}
\label{fig:maxsim}
\end{figure}

The mechanism of multi-vector retriever (e.g., ColBERT) involves identifying the most similar document tokens for each query token. 
While prior research \cite{wang2023reproducibility, liu2024analysis} has explored this in \textit{MonoIR}, its behavior in other language contexts remains unexplored. 
This token-level analysis offers a rationale for a key aspect of \textbf{(RQ3)}: understanding why mixed-language queries can outperform native queries for English document retrieval.

Our analysis compared MaxSim token pair similarity (a query token and its maximal similarity document token) in mixed-language versus native queries.
Figure \ref{fig:maxsim} (left) shows mixed-language queries, by including English terms (e.g., "Intellectual Property Rights" from German "Intellektuelle Eigentumsrechte"), allow these English tokens to form MaxSim pairings (\textcolor{red}{\textbf{$\times$}}) with accurate, higher similarity scores.
Conversely, native queries (right) rely on cross-lingual interpretation of native tokens (e.g., German "Intellekt," "Eigen") to map English concepts. While MaxSim pairings are also identified (\textcolor{red}{\textbf{$\times$}}), this mapping yields weaker similarity for such crucial English concepts.
Thus, intentionally mixing English terms improves MaxSim matching through higher similarity scores for English terms—a key rationale (\textbf{RQ3}) for \textit{MQIR}'s enhanced English retrieval.

\section{Conclusion}


This study addressed the prevalent yet understudied phenomenon of mixed-language querying among bilingual speakers by introducing \textbf{MiLQ}—the first public user-crafted \textit{MQIR} benchmark, validated for both realism and high user preference. Our comprehensive experiments on MiLQ revealed that current IR models exhibit inconsistent performance across diverse query types, highlighting the need for more robust retrieval systems and demonstrating the promising potential of code-switched training data. Finally, we discovered that intentionally mixing English terms into queries serves as an effective strategy for enhancing English document retrieval among bilingual users.

\section{Limitations}

While MiLQ is a valuable first public \textit{MQIR} benchmark, it shares limitations common to the broader multilingual IR field. A key challenge is the test set scale; unlike large monolingual English benchmarks (e.g., MS-MARCO \cite{bajaj2018msmarcohumangenerated}, NQ \cite{kwiatkowski2019natural} with thousands of queries), \textit{CLIR} benchmarks typically comprise only tens to hundreds of queries \cite{asai-etal-2021-xor, lawrie2023hc3, lawrie2023overview, soboroff2023better, adeyemi2024ciral}. This is because creating numerous high-quality multilingual test sets is highly resource-intensive. Larger \textit{MQIR} benchmarks would be beneficial, allowing for more robust methodological comparisons and fostering advancements in the field.

MiLQ currently focuses on English-native language pairs, excluding non-English/non-English combinations; future inclusion of these diverse pairings is desirable. Furthermore, while realistic, MiLQ's user-crafted queries may not capture all code-switching patterns, as these are shaped by individual cultural and linguistic experiences. Broader participant involvement could enrich future datasets with more diverse, authentic patterns. Budgetary constraints also limited MiLQ's initial language and domain scope, suggesting future expansions for wider utility.

These limitations and the need for larger test collections highlight promising future directions. Beyond creating larger \textit{MQIR} benchmarks, key research avenues include expanding linguistic diversity (with non-English/non-English pairs), investigating broader code-switching patterns via more diverse annotators, and leveraging advanced techniques like multilingual LLMs to enhance \textit{MQIR}.

\section*{Ethical Considerations}
\paragraph{Dataset Licensing and Usage}
Our work uses three primary datasets: NeuCLIR22 \cite{lawrie2023overview}, CLEF00-03 \cite{braschler2003clef}, and our newly introduced MiLQ dataset. We have verified the licensing terms for all existing datasets and ensured our usage is consistent with their intended research purposes. The NeuCLIR22 dataset is published by NIST with public access level and is subject to the NIST Open License. The CLEF00-03 data is distributed by ELDA under an End-User Agreement for Evaluation Packages for Research Use, which permits evaluation purposes. Our MiLQ dataset will be distributed by ELDA under a free evaluation license for academic organizations. The dataset is accessible through our code repository. The MiLQ dataset and our findings are intended solely for academic research purposes in multilingual information retrieval. We discourage commercial deployment without further evaluation of potential societal impacts and biases.

\paragraph{Human Annotation}
For our MiLQ dataset, queries were created by bilingual speakers and subsequently validated by a different group of bilingual speakers to ensure quality and reduce bias. All participants in both stages were compensated fairly for their work based on regional wage standards and time estimates for each task. The detailed annotation guidelines and compensation structure for each stage are described in Appendix \ref{sec:appendix_Annotation_Details} and \ref{sec:human_eval_guidelines}. We obtained informed consent from all participants, clearly explaining how their contributions would be used in research and dataset creation.

\paragraph{Potential Risks}
The scope of our study is limited to the news domain and nine languages: English, Swahili, Somali, Finnish, German, French, Chinese, Persian, and Russian. Therefore, our findings may not generalize to other domains, genres, or languages not represented in our evaluation. Our findings may inadvertently favor certain language pairs or retrieval approaches that work better for high-resource languages, potentially contributing to digital language divides. Regarding personal information, we followed the existing privacy protection measures established by NIST and ELDA for the original datasets.

\section*{Acknowledgments}
This research was supported by the MSIT (Ministry of Science and ICT), Korea, under the ITRC(Information Technology Research Center) support program (IITP-2025-RS-2020-II201789, Contribution Rate: 47.5\%) supervised by the IITP (Institute for Information \& Communications Technology Planning \& Evaluation); by the Culture, Sports and Tourism R\&D Program through the Korea Creative Content Agency grant funded by the Ministry of Culture, Sports and Tourism in 2025 (Project Name: Development of an AI-Based Korean Diagnostic System for Efficient Korean Speaking Learning by Foreigners, Project Number: RS-2025-02413038, Contribution Rate: 47.5\%); and by Institute of Information \& communications Technology Planning \& Evaluation (IITP) grant funded
by the Korea government(MSIT) (No.RS-2019-II191906, Artificial Intelligence Graduate School Program(POSTECH), Contribution Rate: 5\%).

The MiLQ dataset is released through the ELRA catalogue under the Evaluation Use - ELRA
EVALUATION license, providing free access to researchers for evaluation purposes. If you use this dataset, please include the following citation in your acknowledgements:
MiLQ: Mixed-Language Query Test Set for Bilingual Web Search - Evaluation Package, ELRA catalogue
(http://catalog.elra.info), ISLRN: 317-005-302-361-6, ELRA ID: ELRA-E0047

The following dataset was used for evaluation in this study:
The CLEF Test Suite for the CLEF 2000-2003 Campaigns – Evaluation Package, ELRA catalogue
(http://catalog.elra.info), ISLRN: 200-586-423-805-2, ELRA ID: ELRA-E0008



\bibliography{custom}

\begin{thebibliography}{54}
\providecommand{\natexlab}[1]{#1}

\bibitem[{Adeyemi et~al.(2024)Adeyemi, Oladipo, Zhang, Alfonso-Hermelo, Rezagholizadeh, Chen, Omotayo, Abdulmumin, Etori, Musa et~al.}]{adeyemi2024ciral}
Mofetoluwa Adeyemi, Akintunde Oladipo, Xinyu Zhang, David Alfonso-Hermelo, Mehdi Rezagholizadeh, Boxing Chen, Abdul-Hakeem Omotayo, Idris Abdulmumin, Naome~A Etori, Toyib~Babatunde Musa, et~al. 2024.
\newblock Ciral: A test collection for clir evaluations in african languages.
\newblock In \emph{Proceedings of the 47th International ACM SIGIR Conference on Research and Development in Information Retrieval}, pages 293--302.

\bibitem[{Andrzejewski(1978)}]{andrzejewski1978development}
Bogumi{\l}~W Andrzejewski. 1978.
\newblock The development of a national orthography in somalia and the modernization of the somali language.
\newblock \emph{Horn of Africa}.

\bibitem[{Andrzejewski(1979)}]{andrzejewski1979language}
BW~Andrzejewski. 1979.
\newblock Language reform in somalia and the modernization of the somali vocabulary.
\newblock \emph{Northeast African Studies}, pages 59--71.

\bibitem[{Asai et~al.(2021)Asai, Kasai, Clark, Lee, Choi, and Hajishirzi}]{asai-etal-2021-xor}
Akari Asai, Jungo Kasai, Jonathan Clark, Kenton Lee, Eunsol Choi, and Hannaneh Hajishirzi. 2021.
\newblock \href {https://doi.org/10.18653/v1/2021.naacl-main.46} {{XOR} {QA}: Cross-lingual open-retrieval question answering}.
\newblock In \emph{Proceedings of the 2021 Conference of the North American Chapter of the Association for Computational Linguistics: Human Language Technologies}, pages 547--564, Online. Association for Computational Linguistics.

\bibitem[{Auer(1999)}]{auer1999codeswitching}
Peter Auer. 1999.
\newblock From codeswitching via language mixing to fused lects: Toward a dynamic typology of bilingual speech.
\newblock \emph{International journal of bilingualism}, 3(4):309--332.

\bibitem[{Auer(2013)}]{auer2013code}
Peter Auer. 2013.
\newblock \emph{Code-switching in conversation: Language, interaction and identity}.
\newblock Routledge.

\bibitem[{Bajaj et~al.(2018)Bajaj, Campos, Craswell, Deng, Gao, Liu, Majumder, McNamara, Mitra, Nguyen, Rosenberg, Song, Stoica, Tiwary, and Wang}]{bajaj2018msmarcohumangenerated}
Payal Bajaj, Daniel Campos, Nick Craswell, Li~Deng, Jianfeng Gao, Xiaodong Liu, Rangan Majumder, Andrew McNamara, Bhaskar Mitra, Tri Nguyen, Mir Rosenberg, Xia Song, Alina Stoica, Saurabh Tiwary, and Tong Wang. 2018.
\newblock \href {https://arxiv.org/abs/1611.09268} {Ms marco: A human generated machine reading comprehension dataset}.
\newblock \emph{Preprint}, arXiv:1611.09268.

\bibitem[{Bawa et~al.(2020)Bawa, Khadpe, Joshi, Bali, and Choudhury}]{bawa2020multilingual}
Anshul Bawa, Pranav Khadpe, Pratik Joshi, Kalika Bali, and Monojit Choudhury. 2020.
\newblock Do multilingual users prefer chat-bots that code-mix? let's nudge and find out!
\newblock \emph{Proceedings of the ACM on Human-Computer Interaction}, 4(CSCW1):1--23.

\bibitem[{Bendersky and Kurland(2008)}]{bendersky2008utilizing}
Michael Bendersky and Oren Kurland. 2008.
\newblock Utilizing passage-based language models for document retrieval.
\newblock In \emph{Advances in Information Retrieval: 30th European Conference on IR Research, ECIR 2008, Glasgow, UK, March 30-April 3, 2008. Proceedings 30}, pages 162--174. Springer.

\bibitem[{Bonab et~al.(2019)Bonab, Allan, and Sitaraman}]{bonab2019simulating}
Hamed Bonab, James Allan, and Ramesh Sitaraman. 2019.
\newblock Simulating clir translation resource scarcity using high-resource languages.
\newblock In \emph{Proceedings of the 2019 ACM SIGIR International Conference on Theory of Information Retrieval}, pages 129--136.

\bibitem[{Braschler(2000)}]{braschler2000clef}
Martin Braschler. 2000.
\newblock Clef 2000—overview of results.
\newblock In \emph{Workshop of the Cross-Language Evaluation Forum for European Languages}, pages 89--101. Springer.

\bibitem[{Braschler(2002{\natexlab{a}})}]{10.1007/3-540-45691-0_2}
Martin Braschler. 2002{\natexlab{a}}.
\newblock Clef 2001 --- overview of results.
\newblock In \emph{Evaluation of Cross-Language Information Retrieval Systems}, pages 9--26, Berlin, Heidelberg. Springer Berlin Heidelberg.

\bibitem[{Braschler(2002{\natexlab{b}})}]{braschler2002clef}
Martin Braschler. 2002{\natexlab{b}}.
\newblock Clef 2002—overview of results.
\newblock In \emph{Workshop of the Cross-Language Evaluation Forum for European Languages}, pages 9--27. Springer.

\bibitem[{Braschler(2003)}]{braschler2003clef}
Martin Braschler. 2003.
\newblock Clef 2003--overview of results.
\newblock In \emph{Workshop of the cross-language evaluation forum for european languages}, pages 44--63. Springer.

\bibitem[{Choi et~al.(2023)Choi, Lee, and Lee}]{choi2023toward}
Yunjae~J Choi, Minha Lee, and Sangsu Lee. 2023.
\newblock Toward a multilingual conversational agent: Challenges and expectations of code-mixing multilingual users.
\newblock In \emph{Proceedings of the 2023 CHI conference on human factors in computing systems}, pages 1--17.

\bibitem[{Conneau et~al.(2017)Conneau, Lample, Ranzato, Denoyer, and J{\'e}gou}]{conneau2017word}
Alexis Conneau, Guillaume Lample, Marc'Aurelio Ranzato, Ludovic Denoyer, and Herv{\'e} J{\'e}gou. 2017.
\newblock Word translation without parallel data.
\newblock \emph{arXiv preprint arXiv:1710.04087}.

\bibitem[{Dai and Callan(2019)}]{dai2019deeper}
Zhuyun Dai and Jamie Callan. 2019.
\newblock Deeper text understanding for ir with contextual neural language modeling.
\newblock In \emph{Proceedings of the 42nd international ACM SIGIR conference on research and development in information retrieval}, pages 985--988.

\bibitem[{Das and Gamb{\"a}ck(2014)}]{das2014identifying}
Amitava Das and Bj{\"o}rn Gamb{\"a}ck. 2014.
\newblock Identifying languages at the word level in code-mixed indian social media text.
\newblock In \emph{Proceedings of the 11th International Conference on Natural Language Processing}, pages 378--387.

\bibitem[{Fu(2017)}]{fu2017query}
Hengyi Fu. 2017.
\newblock Query reformulation patterns of mixed language queries in different search intents.
\newblock In \emph{Proceedings of the 2017 conference on conference human information interaction and retrieval}, pages 249--252.

\bibitem[{Fu(2019)}]{fu2019mixed}
Hengyi Fu. 2019.
\newblock Mixed language queries in online searches: A study of intra-sentential code-switching from a qualitative perspective.
\newblock \emph{Aslib Journal of Information Management}, 71(1):72--89.

\bibitem[{Fung et~al.(1999)Fung, Liu, and Cheung}]{fung1999mixed}
Pascale Fung, Xiaohu Liu, and Chi-Shun Cheung. 1999.
\newblock Mixed language query disambiguation.
\newblock In \emph{Proceedings of the 37th Annual Meeting of the Association for Computational Linguistics}, pages 333--340.

\bibitem[{Gardner-Chloros(2009)}]{gardner2009code}
Penelope Gardner-Chloros. 2009.
\newblock \emph{Code-switching}.
\newblock Cambridge university press.

\bibitem[{Gupta et~al.(2014)Gupta, Bali, Banchs, Choudhury, and Rosso}]{gupta2014query}
Parth Gupta, Kalika Bali, Rafael~E Banchs, Monojit Choudhury, and Paolo Rosso. 2014.
\newblock Query expansion for mixed-script information retrieval.
\newblock In \emph{Proceedings of the 37th international ACM SIGIR conference on Research \& development in information retrieval}, pages 677--686.

\bibitem[{Huang et~al.(2023)Huang, Yu, and Allan}]{huang2023improving}
Zhiqi Huang, Puxuan Yu, and James Allan. 2023.
\newblock Improving cross-lingual information retrieval on low-resource languages via optimal transport distillation.
\newblock In \emph{Proceedings of the Sixteenth ACM International Conference on Web Search and Data Mining}, pages 1048--1056.

\bibitem[{Hurst et~al.(2024)Hurst, Lerer, Goucher, Perelman, Ramesh, Clark, Ostrow, Welihinda, Hayes, Radford et~al.}]{hurst2024gpt}
Aaron Hurst, Adam Lerer, Adam~P Goucher, Adam Perelman, Aditya Ramesh, Aidan Clark, AJ~Ostrow, Akila Welihinda, Alan Hayes, Alec Radford, et~al. 2024.
\newblock Gpt-4o system card.
\newblock \emph{arXiv preprint arXiv:2410.21276}.

\bibitem[{Huzaifah et~al.(2024)Huzaifah, Zheng, Chanpaisit, and Wu}]{huzaifah2024evaluating}
Muhammad Huzaifah, Weihua Zheng, Nattapol Chanpaisit, and Kui Wu. 2024.
\newblock Evaluating code-switching translation with large language models.
\newblock In \emph{Proceedings of the 2024 Joint International Conference on Computational Linguistics, Language Resources and Evaluation (LREC-COLING 2024)}, pages 6381--6394.

\bibitem[{Izacard et~al.()Izacard, Caron, Hosseini, Riedel, Bojanowski, Joulin, and Grave}]{izacardunsupervised}
Gautier Izacard, Mathilde Caron, Lucas Hosseini, Sebastian Riedel, Piotr Bojanowski, Armand Joulin, and Edouard Grave.
\newblock Unsupervised dense information retrieval with contrastive learning.
\newblock \emph{Transactions on Machine Learning Research}.

\bibitem[{Joulin et~al.(2016)Joulin, Grave, Bojanowski, Douze, J{\'e}gou, and Mikolov}]{joulin2016fasttext}
Armand Joulin, Edouard Grave, Piotr Bojanowski, Matthijs Douze, H{\'e}rve J{\'e}gou, and Tomas Mikolov. 2016.
\newblock Fasttext.zip: Compressing text classification models.
\newblock \emph{arXiv preprint arXiv:1612.03651}.

\bibitem[{Kamholz et~al.(2014)Kamholz, Pool, and Colowick}]{kamholz2014panlex}
David Kamholz, Jonathan Pool, and Susan~M Colowick. 2014.
\newblock Panlex: Building a resource for panlingual lexical translation.
\newblock In \emph{LREC}, volume~14, pages 3145--3150.

\bibitem[{Kapchits(2019)}]{kapchits2019somali}
Georgi Kapchits. 2019.
\newblock on the somali temporal lexicon.
\newblock \emph{Bildhaan: An International Journal of Somali Studies}, 19(1):7.

\bibitem[{Khattab and Zaharia(2020)}]{khattab2020colbert}
Omar Khattab and Matei Zaharia. 2020.
\newblock Colbert: Efficient and effective passage search via contextualized late interaction over bert.
\newblock In \emph{Proceedings of the 43rd International ACM SIGIR conference on research and development in Information Retrieval}, pages 39--48.

\bibitem[{Kuwanto et~al.(2024)Kuwanto, Agarwal, Winata, and Wijaya}]{kuwanto2024linguistics}
Garry Kuwanto, Chaitanya Agarwal, Genta~Indra Winata, and Derry~Tanti Wijaya. 2024.
\newblock Linguistics theory meets llm: Code-switched text generation via equivalence constrained large language models.
\newblock \emph{arXiv preprint arXiv:2410.22660}.

\bibitem[{Kwiatkowski et~al.(2019)Kwiatkowski, Palomaki, Redfield, Collins, Parikh, Alberti, Epstein, Polosukhin, Devlin, Lee et~al.}]{kwiatkowski2019natural}
Tom Kwiatkowski, Jennimaria Palomaki, Olivia Redfield, Michael Collins, Ankur Parikh, Chris Alberti, Danielle Epstein, Illia Polosukhin, Jacob Devlin, Kenton Lee, et~al. 2019.
\newblock Natural questions: a benchmark for question answering research.
\newblock \emph{Transactions of the Association for Computational Linguistics}, 7:453--466.

\bibitem[{Lawrie et~al.(2023{\natexlab{a}})Lawrie, MacAvaney, Mayfield, McNamee, Oard, Soldaini, and Yang}]{lawrie2023overview}
Dawn Lawrie, Sean MacAvaney, James Mayfield, Paul McNamee, Douglas~W Oard, Luca Soldaini, and Eugene Yang. 2023{\natexlab{a}}.
\newblock Overview of the trec 2022 neuclir track.
\newblock \emph{arXiv preprint arXiv:2304.12367}.

\bibitem[{Lawrie et~al.(2023{\natexlab{b}})Lawrie, Mayfield, Oard, Yang, Nair, and Galu{\v{s}}{\v{c}}{\'a}kov{\'a}}]{lawrie2023hc3}
Dawn Lawrie, James Mayfield, Douglas~W Oard, Eugene Yang, Suraj Nair, and Petra Galu{\v{s}}{\v{c}}{\'a}kov{\'a}. 2023{\natexlab{b}}.
\newblock Hc3: A suite of test collections for clir evaluation over informal text.
\newblock In \emph{Proceedings of the 46th International ACM SIGIR Conference on Research and Development in Information Retrieval}, pages 2880--2889.

\bibitem[{Lin et~al.(2021)Lin, Ma, Lin, Yang, Pradeep, and Nogueira}]{lin2021pyserini}
Jimmy Lin, Xueguang Ma, Sheng-Chieh Lin, Jheng-Hong Yang, Ronak Pradeep, and Rodrigo Nogueira. 2021.
\newblock Pyserini: An easy-to-use python toolkit to support replicable ir research with sparse and dense representations.
\newblock \emph{arXiv preprint arXiv:2102.10073}.

\bibitem[{Litschko et~al.(2023)Litschko, Artemova, and Plank}]{litschko2023boosting}
Robert Litschko, Ekaterina Artemova, and Barbara Plank. 2023.
\newblock Boosting zero-shot cross-lingual retrieval by training on artificially code-switched data.
\newblock \emph{arXiv preprint arXiv:2305.05295}.

\bibitem[{Litschko et~al.(2025)Litschko, Kraus, Blaschke, and Plank}]{litschko2025cross}
Robert Litschko, Oliver Kraus, Verena Blaschke, and Barbara Plank. 2025.
\newblock Cross-dialect information retrieval: Information access in low-resource and high-variance languages.
\newblock In \emph{Proceedings of the 31st International Conference on Computational Linguistics}, pages 10158--10171.

\bibitem[{Liu et~al.(2025)Liu, Xu, Zhang, and Lin}]{liu2025impact}
Andrew Liu, Edward Xu, Crystina Zhang, and Jimmy Lin. 2025.
\newblock The impact of incidental multilingual text on cross-lingual transfer in monolingual retrieval.
\newblock In \emph{European Conference on Information Retrieval}, pages 165--173. Springer.

\bibitem[{Liu et~al.(2024)Liu, Guo, Mao, Dou, Wen, Jiang, Zhang, and Cao}]{liu2024analysis}
Qi~Liu, Gang Guo, Jiaxin Mao, Zhicheng Dou, Ji-Rong Wen, Hao Jiang, Xinyu Zhang, and Zhao Cao. 2024.
\newblock An analysis on matching mechanisms and token pruning for late-interaction models.
\newblock \emph{ACM Transactions on Information Systems}, 42(5):1--28.

\bibitem[{Myers-Scotton(1997)}]{myers1997duelling}
Carol Myers-Scotton. 1997.
\newblock \emph{Duelling languages: Grammatical structure in codeswitching}.
\newblock Oxford University Press.

\bibitem[{Nair et~al.(2022)Nair, Yang, Lawrie, Duh, McNamee, Murray, Mayfield, and Oard}]{nair2022transfer}
Suraj Nair, Eugene Yang, Dawn Lawrie, Kevin Duh, Paul McNamee, Kenton Murray, James Mayfield, and Douglas~W Oard. 2022.
\newblock Transfer learning approaches for building cross-language dense retrieval models.
\newblock In \emph{European Conference on Information Retrieval}, pages 382--396. Springer.

\bibitem[{Nakatani(2010)}]{nakatani2010langdetect}
Shuyo Nakatani. 2010.
\newblock \href {https://github.com/shuyo/language-detection} {Language detection library for java}.

\bibitem[{Nguyen et~al.(2016)Nguyen, Rosenberg, Song, Gao, Tiwary, Majumder, and Deng}]{nguyen2016ms}
Tri Nguyen, Mir Rosenberg, Xia Song, Jianfeng Gao, Saurabh Tiwary, Rangan Majumder, and Li~Deng. 2016.
\newblock Ms marco: A human-generated machine reading comprehension dataset.

\bibitem[{OpenAI(2023)}]{openai2023chatgpt}
OpenAI. 2023.
\newblock Chatgpt.
\newblock \url{https://chat.openai.com}.
\newblock \url{https://chat.openai.com}.

\bibitem[{Robertson et~al.(2009)Robertson, Zaragoza et~al.}]{robertson2009probabilistic}
Stephen Robertson, Hugo Zaragoza, et~al. 2009.
\newblock The probabilistic relevance framework: Bm25 and beyond.
\newblock \emph{Foundations and Trends{\textregistered} in Information Retrieval}, 3(4):333--389.

\bibitem[{Sequiera et~al.(2015)Sequiera, Choudhury, Gupta, Rosso, Kumar, Banerjee, Naskar, Bandyopadhyay, Chittaranjan, Das et~al.}]{sequiera2015overview}
Royal Sequiera, Monojit Choudhury, Parth Gupta, Paolo Rosso, Shubham Kumar, Somnath Banerjee, Sudip~Kumar Naskar, Sivaji Bandyopadhyay, Gokul Chittaranjan, Amitava Das, et~al. 2015.
\newblock Overview of fire-2015 shared task on mixed script information retrieval.
\newblock In \emph{FIRE workshops}, volume 1587, pages 19--25.

\bibitem[{Sitaram et~al.(2019)Sitaram, Chandu, Rallabandi, and Black}]{sitaram2019survey}
Sunayana Sitaram, Khyathi~Raghavi Chandu, Sai~Krishna Rallabandi, and Alan~W Black. 2019.
\newblock A survey of code-switched speech and language processing.
\newblock \emph{arXiv preprint arXiv:1904.00784}.

\bibitem[{Soboroff(2023)}]{soboroff2023better}
Ian Soboroff. 2023.
\newblock The better cross-language datasets.
\newblock In \emph{Proceedings of the 46th International ACM SIGIR Conference on Research and Development in Information Retrieval}, pages 3047--3053.

\bibitem[{Voorhees(2005)}]{voorhees2005trec}
EM~Voorhees. 2005.
\newblock Trec: Experiment and evaluation in information retrieval.

\bibitem[{Wang et~al.(2023)Wang, Macdonald, Tonellotto, and Ounis}]{wang2023reproducibility}
Xiao Wang, Craig Macdonald, Nicola Tonellotto, and Iadh Ounis. 2023.
\newblock Reproducibility, replicability, and insights into dense multi-representation retrieval models: from colbert to col.
\newblock In \emph{Proceedings of the 46th International ACM SIGIR Conference on Research and Development in Information Retrieval}, pages 2552--2561.

\bibitem[{Winata et~al.(2023)Winata, Aji, Yong, and Solorio}]{winata2023decades}
Genta Winata, Alham~Fikri Aji, Zheng~Xin Yong, and Thamar Solorio. 2023.
\newblock The decades progress on code-switching research in nlp: A systematic survey on trends and challenges.
\newblock In \emph{Findings of the Association for Computational Linguistics: ACL 2023}, pages 2936--2978.

\bibitem[{Yang et~al.(2024)Yang, Lawrie, Mayfield, Oard, and Miller}]{yang2024translate}
Eugene Yang, Dawn Lawrie, James Mayfield, Douglas~W Oard, and Scott Miller. 2024.
\newblock Translate-distill: Learning cross-language dense retrieval by translation and distillation.
\newblock In \emph{European Conference on Information Retrieval}, pages 50--65. Springer.

\bibitem[{Yong et~al.(2023)Yong, Zhang, Forde, Wang, Subramonian, Lovenia, Cahyawijaya, Winata, Sutawika, Cruz et~al.}]{yong2023prompting}
Zheng~Xin Yong, Ruochen Zhang, Jessica Forde, Skyler Wang, Arjun Subramonian, Holy Lovenia, Samuel Cahyawijaya, Genta Winata, Lintang Sutawika, Jan Christian~Blaise Cruz, et~al. 2023.
\newblock Prompting multilingual large language models to generate code-mixed texts: The case of south east asian languages.
\newblock In \emph{Proceedings of the 6th Workshop on Computational Approaches to Linguistic Code-Switching}, pages 43--63.

\end{thebibliography}
\clearpage
\appendix
\onecolumn

\section{Data Annotation}

\subsection{\label{sec:appendix_Annotation_Details} Details of the Employment and Annotation}

\begin{figure*}[h]
\centering
\includegraphics[width=1.0\textwidth]{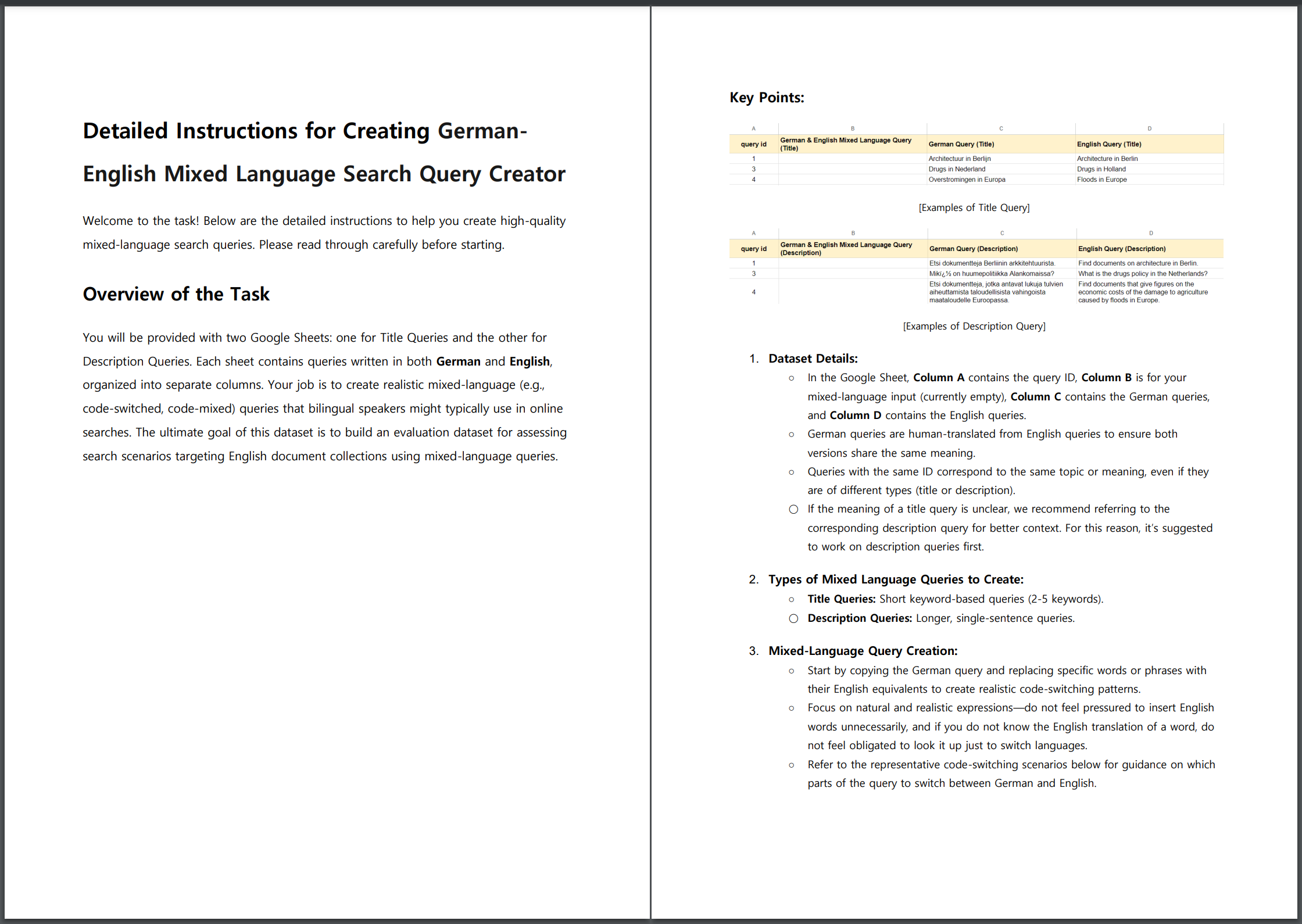} 
\caption{Guideline for German-English mixed-language search query annotators.}
\label{fig:guideline}
\end{figure*}

We recruited bilingual speakers through Upwork\footnote{https://www.upwork.com}, who were fluent in both English and one of the following languages: Swahili (SW), Somali (SO), Finnish (FI), German (DE), French (FR), Chinese (ZH), Persian (FA), or Russian (RU).
These annotators were selected based on their proficiency in both languages and their extensive experience in translation activities between English and their respective languages.
We provided the annotators with clear guidelines, as shown in Figure \ref{fig:guideline}.
The payment was based on the number of queries, with SW, SO, FI, DE, and FR totaling 302 queries (Title + Description) for \$40. For ZH, FA, and RU, we created 94, 90, and 88 queries, respectively, with a total cost of \$20 per language.

\newpage

\subsection{\label{sec:appendix_title_example} Examples of Title and Description queries in MiLQ}

This appendix illustrates Title and Description mixed-language queries (MiLQ) from our dataset, derived from native and English sources. The figures highlight code-switched segments and indicate their Code-Mixing Index (CMI), calculated by \ref{eq:CMI}.

\begin{figure*}[h]
\centering
\includegraphics[width=1.0\textwidth]{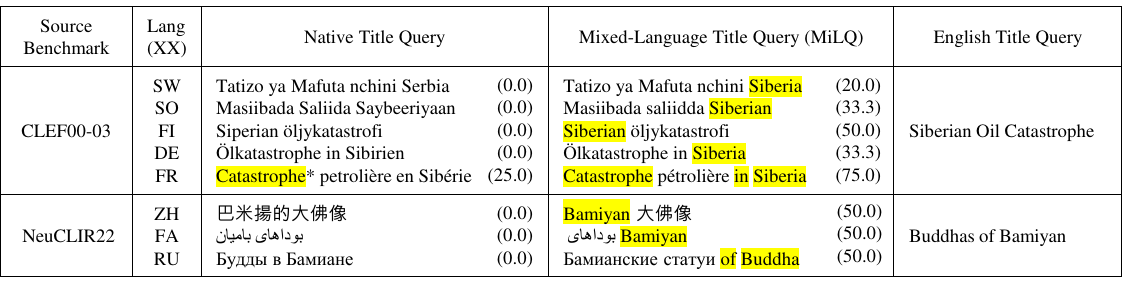} 
\caption{Examples of Title queries from the MiLQ dataset. Code-switched segments are highlighted, and CMI values are shown in parentheses. (*Note: Although 'Catastrophe' is also a French word, it was identified as English by the language model in this instance.)}
\label{fig:milq_examples_title}
\end{figure*}

\begin{figure*}[ht!]
\centering
\includegraphics[width=1.0\textwidth]{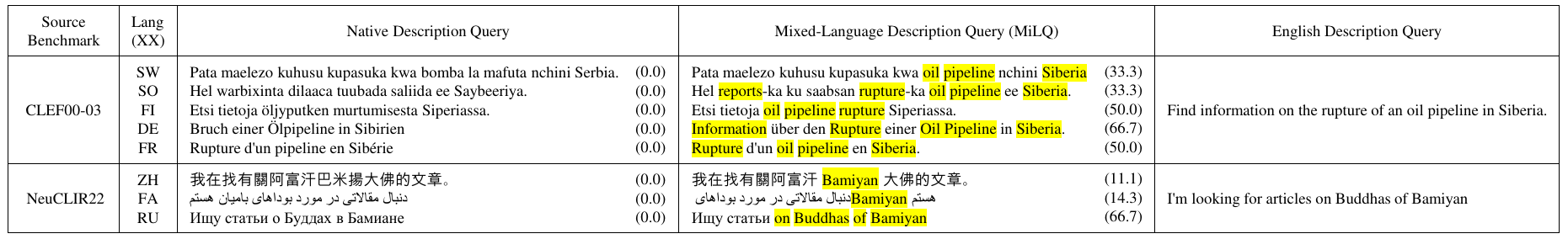} 
\caption{Examples of Description queries from the MiLQ dataset, corresponding to the same query IDs as the Title examples shown in Figure \ref{fig:milq_examples_title}. Code-switched segments are highlighted, and CMI values are indicated in parentheses.}
\label{fig:milq_examples_description}
\end{figure*}

\subsection{\label{sec:cmi} Code-Mixing Index (CMI)}

\begin{figure*}[h!]
\centering
\includegraphics[width=1.0\textwidth]{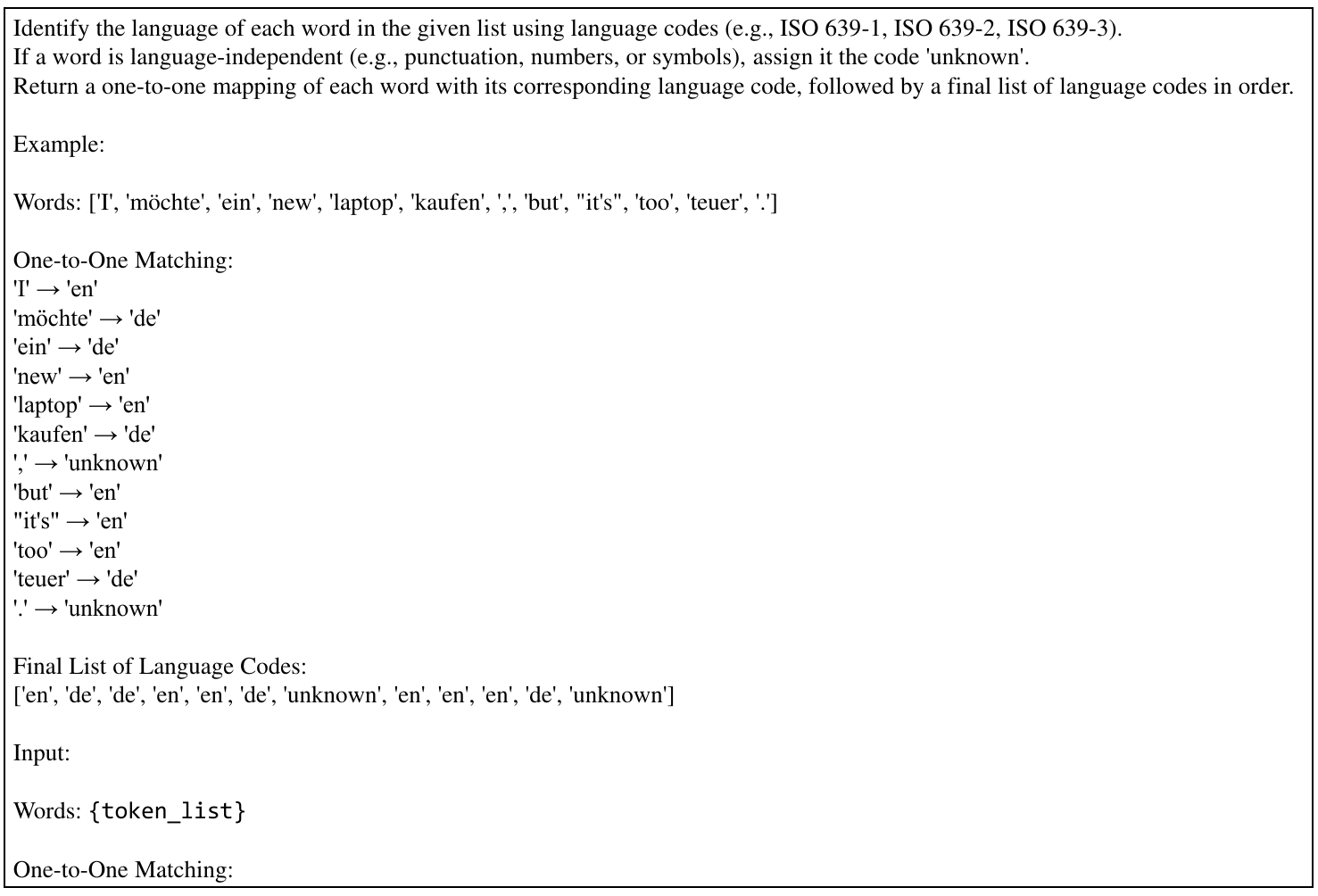} 
\caption{Prompt template for language identification.}
\label{fig:langDetect}
\end{figure*}

The formula for the Code-Mixing Index (CMI) \cite{das2014identifying} is as follows:

\begin{equation}
CMI =
\begin{cases}
100 \times \left( 1 - \frac{\max(w_i)}{n - u} \right) & \text{if } n > u \\
0 & \text{if } n = u,
\end{cases}
\label{eq:CMI}
\end{equation}
where \(w_i\) is the word count in language \(i\), \(\max(w_i)\) is the word count in the primary language, \(n\) is the total word count, and \(u\) is the number of language-independent tokens (e.g., numbers, hashtags).
In our analysis, we treat the primary language as the native language.
We used GPT-4o \cite{hurst2024gpt} instead of existing tools for more precise language identification. While existing tools such as language-detection \cite{nakatani2010langdetect} and fastText \cite{joulin2016fasttext} have been widely used for language identification, we observed certain inconsistencies in accuracy. 
Therefore, we leveraged LLMs for more accurate data analysis. 
First, we tokenize the text at the word level using NLTK\footnote{https://www.nltk.org/}. 
For Chinese text, we apply Jieba\footnote{https://github.com/fxsjy/jieba}, a specialized tokenizer optimized for Chinese word segmentation. 
After tokenization, we utilize GPT-4o to classify each token's language using the prompt template shown in Figure \ref{fig:langDetect}.

\newpage

\subsection{\label{sec:gpt_eval_rubrics} GPT-Evaluation Rubric}

For GPT-based evaluation, we adopted the Accuracy and Fluency rubrics from \citeauthor{kuwanto2024linguistics}, using their publicly available prompts and evaluation code framework.\footnote{\url{https://github.com/gkuwanto/ezswitch}} While their assessments utilized GPT-4O-mini, our study employed the more powerful GPT-4o. The model was instructed to evaluate generated code-switched sentences against the original monolingual sentences on a 1 (lowest) to 3 (highest) scale for each criterion.

\paragraph{Accuracy}
This criterion measures how well the generated sentence preserves the meaning and information of the original sentence, and whether the code-switched terms are used correctly and appropriately.
\begin{description}
\itemsep0em
\item[Score 1 (Low):] Significant deviation from original meaning; key information missing, altered, or redundantly repeated. Code-switched terms incorrect/inappropriate. Introduces new information.
\item[Score 2 (Moderate):] Minor deviation from original meaning; most key information present but may have slight errors. Most code-switched terms appropriate with minor mistakes.
\item[Score 3 (High):] Fully preserves original meaning; all key information present and correct. Code-switched terms accurate and appropriately used.
\end{description}

\paragraph{Fluency}
This criterion measures how natural and easy to understand the generated sentence is, considering grammar, syntax, and the smooth integration of code-switching.
\begin{description}
\itemsep0em
\item[Score 1 (Low):] Sentence is difficult to understand or awkward; poor grammar/syntax in either language. Code-switching disrupts sentence flow.
\item[Score 2 (Moderate):] Sentence is understandable but may have awkward/unnatural phrasing; acceptable grammar/syntax. Code-switching somewhat smooth but not perfectly integrated.
\item[Score 3 (High):] Sentence is natural and easy to understand; good grammar/syntax in both languages. Code-switching is smooth and seamless, enhancing flow.
\end{description}

\subsection{\label{sec:human_eval_guidelines} Human-Evaluation Guidelines and Rubrics}

\begin{figure*}[htbp]
\centering
\fbox{\includegraphics[width=0.83\textwidth]{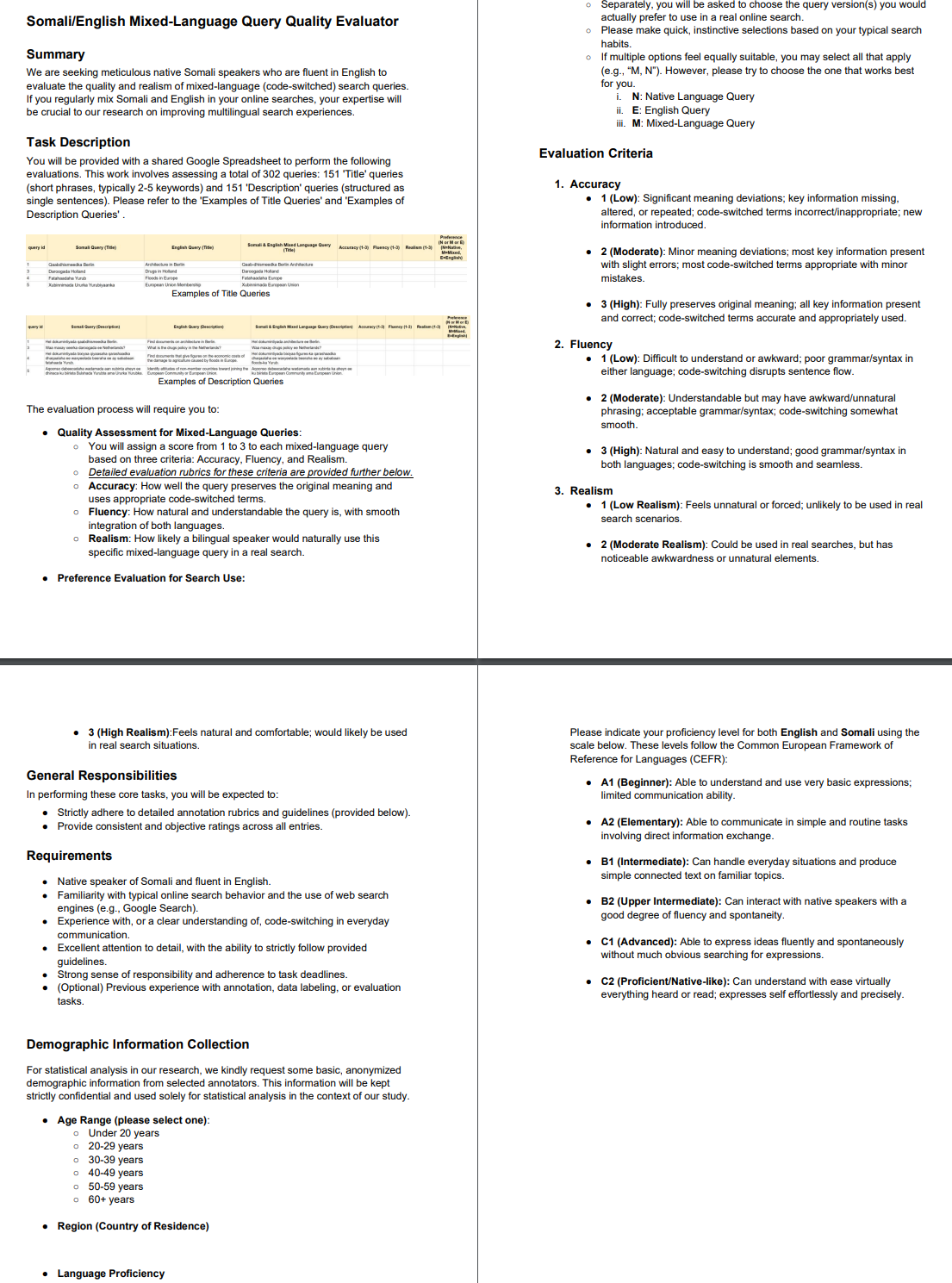}}
\caption{An example of the detailed annotation guidelines provided to bilingual evaluators, in this case for Somali-English mixed-language search queries. Similar guidelines were adapted for other language pairs.}
\label{fig:human_eval_guideline_example} 
\end{figure*}

For the human evaluation of mixed-language queries (MiLQ), we again recruited bilingual speakers via Upwork. Eligibility required proficiency in English and one target language at least at the B2 CEFR level, plus prior translation or linguistic experience, ensuring high-quality judgments. Annotators received detailed instructions (see Figure \ref{fig:human_eval_guideline_example}) and evaluated MiLQ quality using three criteria: Accuracy, Fluency, and Realism, rated on a 1-3 scale.

The payment scheme for this evaluation reflected task complexity and language availability: SO and SW annotators were compensated at \$20 per annotator; FI, FR, and DE annotators at \$30; and FA, ZH, and RU annotators at \$15 each.

\subsubsection*{Accuracy and Fluency Rubrics}

Accuracy and Fluency rubrics mirrored those used in GPT-Evaluation (see Appendix \ref{sec:gpt_eval_rubrics}). Accuracy measures how well a MiLQ preserves the original query’s meaning and appropriately integrates code-switched terms. Fluency assesses the naturalness and clarity of language mixing, ensuring smooth integration of both languages.

\paragraph{Realism}
This criterion, specific to human evaluation, assesses the likelihood that a bilingual speaker would naturally produce or use the given MiLQ in a real online search context.
\begin{description}
\itemsep0em
\item[Score 1 (Low):] Query feels unnatural or forced; unlikely to be used in real search scenarios.
\item[Score 2 (Moderate):] Query could be used in real searches, but has noticeable awkwardness or unnatural elements.
\item[Score 3 (High):] Query feels natural and comfortable; would likely be used in real search situations.
\end{description}

\newpage

\subsection{\label{sec:appendix_preference_discussion} Insights from Annotator Interviews on Mixed-Language Query Usage}

To gain deeper insights into why and when bilingual users employ mixed-language queries (MiLQ) in real-world online searches, we conducted semi-structured interviews with all annotators. 
A common question posed was: "\textit{In what situations are mixed-language queries commonly used in real-world online search contexts, and for what reasons?}" Table \ref{tab:interview_summary} summarizes the key themes derived from their responses.

\begin{table*}[htbp]
    \centering
    \resizebox{\textwidth}{!}{%
    \begin{tabular}{p{0.15\textwidth} p{0.63\textwidth} p{0.17\textwidth}} 
    \toprule
    \textbf{Key Theme} & \textbf{Description: Why Users Mix Languages in Queries} & \textbf{Relevant Languages Mentioned} \\
    \midrule
    \textbf{Lexical Gaps} &
    Native language lacks suitable or clear terms for specific concepts (esp. modern/technical), or English equivalents offer greater precision, familiarity, or avoid awkward/ambiguous translations. &
    Swahili, Somali, French, Finnish, German, Chinese, Persian, Russian \\
    \midrule
    \textbf{Broader Information Access} &
    English terms are used to retrieve a broader range or larger volume of online results, especially when native-language queries are perceived as too restrictive or yield insufficient/biased information. & 
    Swahili, Somali, Finnish, German, Chinese, Persian \\
    \midrule
    \textbf{Querying Efficiency} &
    English terms are preferred for faster query formulation, due to shorter terms, keyboard convenience, or greater cognitive accessibility (i.e., English terms come to mind more readily or are more familiar).  &
    Swahili, Somali, French, Finnish, German, Chinese, Russian  \\
    \midrule
    \textbf{Grammatical / Orthographic Simplification} &
    English terms are selected to bypass complex native grammatical constructions (e.g., inflections, case agreement for foreign words) or challenges posed by keyboard layouts and non-Latin scripts.   &
    Russian \\
    \midrule
    \textbf{Language Modernization} &
    Reliance on established English terms arises from insufficient institutional efforts to standardize native terminology for contemporary concepts, especially in digital and tech domains  &
    Somali \\
    \bottomrule
    \end{tabular}%
    }
    \caption{Summary of Key Motivations for Mixed-Language Query Usage (Condensed to 5 Points) from Annotator Interviews}
    \label{tab:interview_summary} 
\end{table*}

In essence, these interviews highlight that bilinguals employ MiLQ for diverse, practical reasons. Key drivers include bridging lexical gaps or seeking terminological precision when native terms are inadequate, especially for modern or technical concepts. Users also mix languages to expand information access, retrieving broader or more diverse results than native-only queries might yield, or to overcome perceived biases. Querying efficiency and fluency are other significant factors, with English often offering faster input or more readily accessible terms. Furthermore, mixed-language can serve to simplify grammatical or orthographic complexities inherent in some native languages, or address deficiencies in language modernization where native terminology for contemporary concepts is lacking.

It is important to note that the specific motivations and patterns of Mixed-language query usage are often highly speaker- and context-dependent, influenced by individual linguistic backgrounds, cognitive habits, the nature of the information need, and even momentary contextual factors. Understanding these varied drivers is crucial for developing IR systems that can effectively cater to the nuanced and dynamic search behaviors of bilingual users worldwide.

\newpage

\subsection{\label{sec:appendix_mixed_pos} Part-of-Speech Distribution of Code-Switched Words in Queries}

\begin{figure*}[h]
\centering
\includegraphics[width=0.9\textwidth]{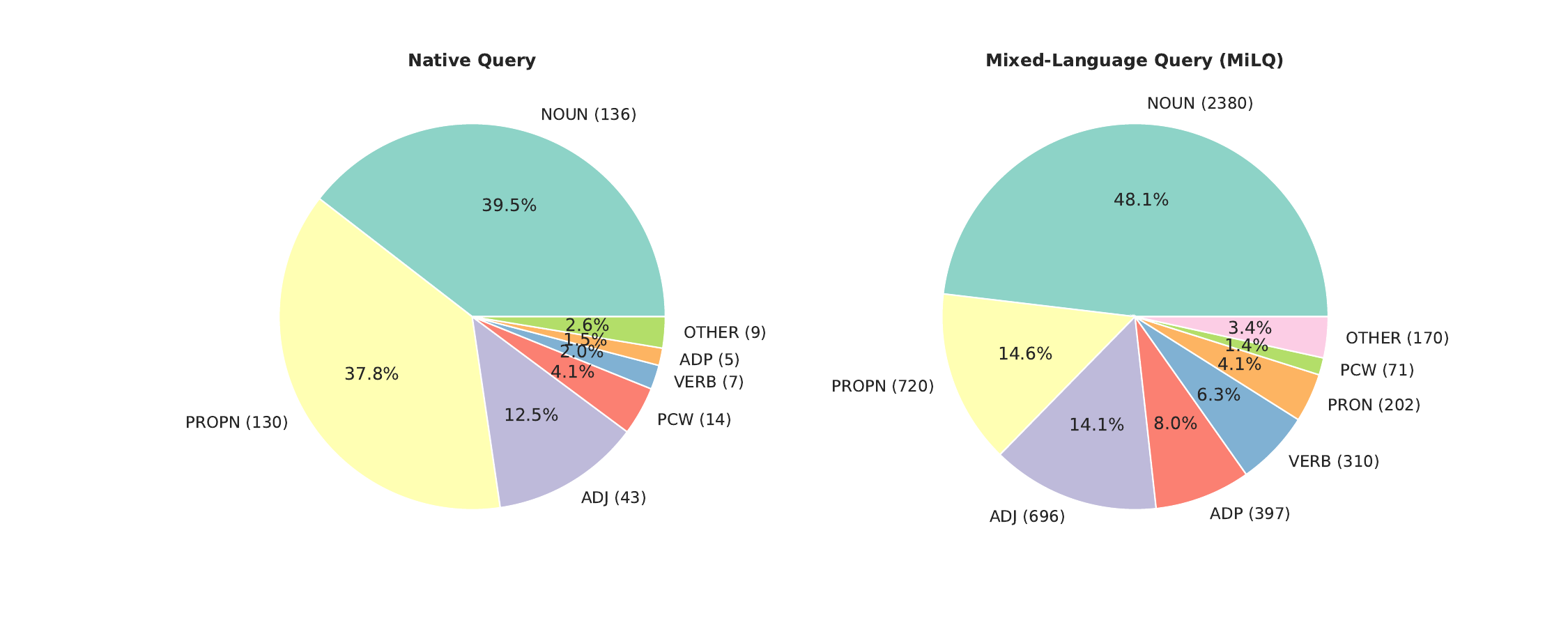} 
\caption{POS distribution of English code-switched words in queries from NeuCLIR22 and CLEF00-03 (left) and MiLQ dataset (right). PCW refers to punctuation-combined words.}
\label{fig:pos}
\end{figure*}

The distribution of English words in both native and mixed-language queries predominantly shows that nouns and proper nouns are the most common parts of speech. However, in our MiLQ dataset, nouns outnumber proper nouns, which contrasts with the distribution observed in native queries. Moreover, our dataset exhibits code-switching not only in nouns and proper nouns but also in a broader range of parts of speech, including adjectives, prepositions, verbs, and pronouns, showing a more diverse pattern of code-switching compared to existing datasets.

\section{Experiment Details}

\subsection{\label{sec:appendixA.2} Benchmark Statistics}

\begin{table*}[ht]
\footnotesize
\centering
\begin{tabular}{l|ccc|cccccc}
\toprule
 & \multicolumn{3}{c|}{NeuCLIR 22} & \multicolumn{6}{c}{CLEF00-03} \\
 & ZH & FA & RU & SW & SO & FI & DE & FR & EN \\ \midrule
\# Queries & 47 & 45 & 44 & 151 & 151 & 151 & 151 & 151 & 151 \\ \midrule
\# Documents & 3.2M & 2.2M & 4.6M & -- & -- & -- & -- & -- & 113k \\
\# Passages & 19.8M & 14.0M & 25.1M & -- & -- & -- & -- & -- & 1.01M\\ \bottomrule
\end{tabular}%
\caption{NeuCLIR22 and CLEF00-03 benchmark statistics.}
\label{table:collection_stats}
\end{table*}

Following previous research \cite{huang2023improving}, we use 151 queries from the CLEF C001 – C200 topics, excluding those with no relevant judgments. English documents are sourced from the Los Angeles Times corpus, which includes 113k news articles.
For high-resource languages such as Finnish, German, and French, queries are directly provided by the CLEF campaign. In contrast, for low-resource languages, \citeauthor{bonab2019simulating} provided Somali and Swahili translations of English queries.

\subsection{\label{sec:metric} Evaluation Metrics}

We evaluate retrieval performance using two standard Information Retrieval metrics:
\textbf{MAP@100 (Mean Average Precision at 100)}: Evaluates ranked lists by averaging precision scores after each relevant binary-judged document is retrieved, up to 100 results. Higher scores indicate better overall retrieval.
\textbf{nDCG@20 (normalized Discounted Cumulative Gain at 20)}: Assesses ranked lists by measuring cumulative gain from graded-relevance documents within the top 20, discounted by rank and normalized by the ideal gain. Higher scores mean better top-ranking of highly relevant items.

\subsection{\label{sec:appendix_implementation_details} Implementation Details}

\paragraph{Model Configuration}

For BM25, we utilized the Pyserini toolkit \cite{lin2021pyserini}, which provides reproducible sparse retrieval through Lucene \footnote{\url{https://github.com/castorini/pyserini}}. Dense retrieval experiments with mContriever \cite{izacardunsupervised} employed \texttt{facebook/mcontriever-msmarco} model\footnote{\url{https://huggingface.co/facebook/mcontriever-msmarco}}.

Our primary retrieval experiments use the ColBERT architecture \cite{khattab2020colbert}, a multi-vector approach for dense retrieval. We utilized the publicly available PLAID-X implementation\footnote{\url{https://github.com/hltcoe/ColBERT-X}} for all model training and inference. Consistent with standard ColBERT practices, most training artifacts and hyperparameters were adopted directly. Our primary modification involved setting the maximum document passage length to 180 tokens. 
Following established methods \cite{bendersky2008utilizing, dai2019deeper}, documents longer than this threshold were segmented into 180-token passages. During evaluation, the score for each document was determined using the maximum passage score (MaxP) strategy.

All experiments were conducted with a single run. Due to the approximate nearest neighbor (ANN) search employed in ColBERT, experimental results may exhibit minor variations depending on the indexing process. However, we observed that such variations do not lead to substantial differences.

\paragraph{Model Backbones and Computational Resources}
We fine-tuned distinct ColBERT models for each source benchmark dataset, selecting multilingual Pre-trained Language Model (mPLM) backbones based on practices in prior relevant research \cite{yang2024translate, huang2023improving}.

\begin{itemize}
    \item \textbf{For NeuCLIR22 (ZH, FA, RU):} ColBERT was initialized using the XLM-RoBERTa Large model\footnote{\url{https://huggingface.co/FacebookAI/xlm-roberta-large}}. This model contains approximately 561 million parameters. Fine-tuning for these languages was conducted about 48 hours.
   \item \textbf{For CLEF00-03 (SW, SO, FI, DE, FR):} mBERT-base-uncased\footnote{\url{https://huggingface.co/google-bert/bert-base-multilingual-uncased}} was employed as the ColBERT encoder. This model has approximately 179 million parameters. Fine-tuning for these languages took about 24 hours.
\end{itemize}
All models were trained on a system equipped with four NVIDIA A100-80GB GPUs. During the training process, model is trained with 6 passages for each query.

\paragraph{Hyperparameters}
A common set of optimization hyperparameters was used for fine-tuning all models. We employed the AdamW optimizer with a learning rate of 5e-6. All models underwent training for 200,000 steps. The total effective batch size was 64, achieved by using a batch size of 16 per GPU across the four GPUs.

\newpage

\subsection{\label{sec:appendix_individual_performance}Performance in Individual Languages}

\begin{table*}[h]
    \centering
    \small
    \renewcommand{\arraystretch}{0.75}
    \begin{tabular}{l | ccccccc}
        \toprule
        \small{Query Lang} & \small{BM25} & \small{Mono-Distil} & \small{Mixed-Distil} & \small{Cross-Distil} & \small{mContriever} & \scriptsize{NMT$\rightarrow$BM25} & \scriptsize{NMT$\rightarrow$Mono-Distil} \\
        \midrule
        SW     & 11.56  & 15.93  & 26.64  & 35.56  & 24.52  & \underline{43.70}  & \textbf{51.00}  \\
        SW\&EN  & 35.82  & 38.21  & \underline{48.04}  & 47.77  & 32.27  & 47.76  & \textbf{56.15}  \\
        EN     & 48.71  & \textbf{57.93}  & 56.53  & 50.77  & 34.88  & 48.71  & \textbf{57.93}  \\
        \midrule
        SO     & 13.14  & 3.12   & 11.07  & 24.91  & 5.58   & \underline{38.43}  & \textbf{49.28}  \\
        SO\&EN  & 40.88  & 34.46  & 43.69  & \underline{49.77}  & 25.18  & 48.44  & \textbf{57.69}  \\
        EN     & 48.71  & \textbf{57.93}  & 56.72  & 49.94  & 34.88  & 48.71  & \textbf{57.93}  \\
        \midrule
        FI     & 7.02   & 29.65  & 40.50  & 41.99  & 27.23  & \underline{45.06}  & \textbf{55.14}  \\
        FI\&EN  & 40.59  & 45.87  & \underline{49.94}  & 49.07  & 32.23  & 45.50  & \textbf{56.20}  \\
        EN     & 48.71  & \textbf{57.93}  & 55.73  & 51.53  & 34.88  & 48.71  & \textbf{57.93}  \\
        \midrule
        DE     & 11.70  & 44.84  & 47.16  & \underline{49.69}  & 30.15  & 45.70  & \textbf{56.60}  \\
        DE\&EN  & 42.33  & \underline{56.89}  & 55.43  & 54.99  & 32.06  & 47.58  & \textbf{57.51}  \\
        EN     & 48.71  & \textbf{57.93}  & 56.70  & 52.12  & 34.88  & 48.71  & \textbf{57.93}  \\
        \midrule
        FR     & 6.95   & 49.46  & 51.96  & \underline{54.95}  & 32.23  & 47.27  & \textbf{57.02}  \\
        FR\&EN  & 21.84  & 54.02  & \underline{55.66}  & 55.59  & 33.53  & 48.17  & \textbf{56.64}  \\
        EN     & 48.71  & \textbf{57.93}  & 57.35  & 53.05  & 34.88  & 48.71  & \textbf{57.93}  \\
        \midrule
        \textit{CLIR}   & 10.07  & 28.60  & 35.41  & 41.42  & 23.94  & \underline{44.03}  & \textbf{53.81}  \\
        \textit{MQIR}   & 36.29  & 45.89  & \underline{50.55}  & 51.44  & 31.05  & 47.49  & \textbf{56.84}  \\
        \textit{MonoIR} & 48.71  & \textbf{57.93}  & 56.60  & 51.48  & 34.88  & 48.71  & \textbf{57.93}  \\
        \bottomrule
    \end{tabular}
    \caption{Performance comparison of different retrieval models across multiple language settings for retrieving English documents. This table presents the performance of individual query languages in this scenario. Additionally, XX\&EN represents queries mixing the native language and English. The metric used is MAP@100 (\%). The best score(s) for each individual language query type (row) are indicated in \textbf{bold}. If there is a unique best score, the second best score(s) are \underline{underlined}.}
    \label{tab:performance_comparison_full1}
\end{table*}

\begin{table*}[h]
    \centering
    \small
    \renewcommand{\arraystretch}{0.9} 
    \begin{tabular}{l c c c c c}
        \toprule
        \textbf{Query Lang} & \textbf{BM25} & \textbf{Mono-Distill} & \textbf{Mixed-Distill} & \textbf{Cross-Distill} & \textbf{mContriever} \\
        \midrule
        ZH        & 25.72  & 46.82  & \textbf{49.59}  & \underline{48.61}  & 32.90  \\
        ZH \& EN  & 3.67   & 41.13  & \textbf{47.46}  & \underline{45.48}  & 19.54  \\
        EN        & 5.74   & 38.52  & \underline{48.73}  & \textbf{48.91}  & 21.22  \\
        \midrule
        FA        & 34.29  & \textbf{48.97}  & 46.06  & \underline{47.26}  & 15.26  \\
        FA \& EN  & 26.02  & \textbf{48.69}  & 45.36  & \underline{45.39}  & 12.97  \\
        EN        & 0.07   & 46.29  & \underline{47.28}  & \textbf{47.99}  & 11.69  \\
        \midrule
        RU        & 36.56  & 47.97  & \underline{48.95}  & \textbf{49.46}  & 36.86  \\
        RU \& EN  & 6.14   & 44.99  & \textbf{49.71}  & \underline{48.02}  & 32.74  \\
        EN        & 1.11   & 44.66  & \textbf{51.42}  & \underline{51.30}  & 29.27  \\
        \midrule
        \textit{MonoIR}      & 32.19  & 47.92  & \underline{48.20}  & \textbf{48.44}  & 28.34  \\
        \textit{MQIR}      & 11.94  & 44.94  & \textbf{47.51}  & \underline{46.30}  & 21.75  \\
        \textit{CLIR}    & 2.31   & 43.16  & \underline{49.14}  & \textbf{49.40}  & 20.73  \\
        \bottomrule
    \end{tabular}
    \caption{Performance comparison of different retrieval models across multiple language settings for retrieving the native documents. This table presents the performance of individual query languages in this scenario. Additionally, XX\&EN represents queries mixing the native language and English. The metric used is nDCG@20 (\%). The best score(s) for each individual language query type (row) are indicated in \textbf{bold}. If there is a unique best score, the second best score(s) are \underline{underlined}.}
    \label{tab:performance_comparison_full2}
\end{table*}

\end{document}